\documentclass[11pt, oneside]{amsart}   	
\usepackage{geometry}                		
\usepackage[utf8]{inputenc}
\usepackage{graphicx}				
\usepackage{epsfig}		
\usepackage{amssymb}
\newcommand{\RR}{\mathbb{R}}
\newcommand{\mcl}{\mathcal}
\numberwithin{equation}{section}

\title{Statistical Methods for the meta-analysis paper by Itzhaky \emph{et al}} 
\author{Steven P. Ellis}
\thanks{Acknowledgements are given at the end.}
\date{December 9, 2021}							

\begin{document}

 \begin{abstract}
This document describes the statistical methods used in Itzhaky \emph{et al}  \cite{lIetAl2021.SuicidePreventionMetaAnalysis}. That paper is a meta-analysis of randomized controlled clinical trials testing methods for preventing suicidal behavior and/or ideation in adolescents. Particularly with respect to self harm behavior, the meta-data are challenging to analyze.

\emph{This} paper has two parts. The first is an informal discussion of the statistical methods used. The second gives detailed mathematical derivations of some formulas and methods. 
 \end{abstract}

\maketitle

\section{Basic Concepts} \label{S:basic.concepts}

This section is based on a Power Point presentation. That's why it's so choppy.

\subsection{Prime Directive: Meta-analyze these papers!} \label{SS:PrimeDirective}
One might argue against the Prime Directive saying that the studies are too heterogeneous to be analyzed as a group. By putting them together we can compare them, but putting them together requires that we should put them on a common footing as much as possible.

We only considered randomized clinical trials and in our summaries of each paper we compared a proposed treatment with a control treatment. Some studies considered more then one proposed treatment. We compared each proposed treatment to a control group. In this way, a paper might be treated as multiple papers. We explored taking this into account later in the meta analysis, but didn't discuss this in the published paper. Suicidality before treatment and during treatment are both reported. (Some studies reported outcomes for different treatment periods. We took the period closest to six months.) 

\subsection{Suicidal Behavior}
The outcomes studied in these papers roughly break down into ideation and behavior. 
Behavior is harder to analyze so \emph{behavior will be the focus of this section.} However, we intentionally made our analysis of ideation follow a similar pattern to the analysis of behavior. Where possible the basic reasoning we present in this section for behavior applies equally well to ideation.

Self Harm Behavior (SHB) as discussed in these papers includes suicide attempt, defined in various ways, and
various forms of self-injurious behavior. We uses attempt whenever possible, but by the Prime Directive we have to accept both. A child who experienced SHB is a ``victim''. 

There are two stages of analyzing SHB:
  \begin{enumerate}
	\item Compute for each paper, a ``figure of merit'' (FM) that summarizes how well the experimental intervention worked compared to the control intervention. Also estimate a standard error (SE) for the FM.  \label{I:compute.FM}
	\item Analyze the collection of FM's and SE's. The main analyses are fixed effects generalized least squares regressions that use the SE's to help model the random component of the model. This allows us to say something about the heterogeneity of the papers. As secondary analyses other meta-data are analyzed.
  \end{enumerate} 
Stage \ref{I:compute.FM} is far more difficult and is the subject of this paper.
Along the way I'll talk about some general statistical concepts that are broadly useful.
\emph{Until the end of this section I discuss the analysis of a single generic study} 

\subsection{KISS (``Keep It Simple, Stupid'')}
``A theory should be as simple as possible, but not simpler.''  A. Einstein said that?
A master applied statistician has the knack for choosing a model as simple possible, but not simpler.

\subsection{Lowest common denominator}
In meta-analysis one has to use very simple models because:
\begin{itemize}
\item One doesn't have subject level data.
\item Only information available in every paper can be included in the model, the lowest common denominator.
\end{itemize}

\subsection{Follow-up}
In most of the papers the ''post'' intervention outcome isn't really post. Many of the interventions go on for some time and usually any event taking place after the beginning of the intervention is counted in the follow-up total. 

\subsection{Behavior outcome summaries}
In practically all the studies, the SHB outcomes are reported as proportion of subjects who exhibited SHB over some time interval. Generically denote such proportions as ``$q$''.
A more informative way of summing up the behavior would be the average number of SHBs per year made by the subjects in the study but few studies present their data that way.
So to put the studies on a common footing we take the least common denominator and work with proportions (of sample or population) who engage in SHB over a specified times interval.

\subsection{Figures of Merit}
To justify our choice of figure of merit (FM), we present a series of FMs, each better than the last.
Our FMs will be based on differences among proportions, but in addition we also use the quotient of the follow-up proportions, the relative risk (subsection \ref{SS:use.prob.models}). Our first attempt is
    \begin{multline*}
       FM_{1} = \text{Improvement in experimental group} - \text{Improvement in control group} \\
          = (q_{flup,exp} - q_{base,exp}) - (q_{flup,con} - q_{base,con}).
    \end{multline*}
Here, ``$flup$'' means ``follow-up'', ``$base$'' means ``baseline''. ``$exp$'' means ``experimental group'', ``$con$'' means ``control group''.

But these studies are RCT's so the two groups are, on average, the same at baseline. 
So a better (less noisy) estimate of the baseline proportion of victims is the proportion, $q_{base}$, in the combined sample. This leads to
    \begin{equation*}
       FM_{2} = (q_{flup,exp} - q_{base}) - (q_{flup,con} - q_{base})
         = q_{flup,exp} - q_{flup,con} .
    \end{equation*}
(I promise I'll later find a use for $q_{base}$.)

\subsection{Follow-up Length Dependence}
A $q$ is the proportion of subjects exhibiting SHB. Exhibiting the behavior within a period of what length? 
Call the time period in which we check for the presence of SHB in a child the ``SHB window.''
Thus, $FM_{2}$ depends on the length of the follow-up SHB window.
Different studies have different follow-up SHB windows.
Therefore, 
    \begin{equation*}
      FM_{1} \text{ and } FM_{2} \text{  are \emph{prima facie} not comparable across studies.}
    \end{equation*}
\emph{This is a fundamental point!}

But the Prime Directive (subsection \ref{SS:PrimeDirective}) requires us to put different studies on a common footing. We accomplish this by ``annualizing''. Use upper case $Q$ to denote proportions (or probabilities) of SHB over a specified one year period. This leads us to a third $FM$:
    \begin{equation*}
      FM_{3} = Q_{flup,exp} - Q_{flup.con} .
    \end{equation*}

Unfortunately, the studies don't provide the ``$Q$'s'', the annual proportions. How can one annualize. 
Just multiply? Suppose 10\% of the subjects engaged in SHB in a 6 month period. Doesn't that mean that 2 x 10\% did over 1 year? Unlikely. Try that with 60\% instead of 10\% and you see that just multiplying will not work. (If we were given the average number of SHB's per subject over a 6 month period we could just muliply by 2, but multiplication doesn't work for probabilities.)

So knowing the proportion for one SHB window, how do you translate that into the proportion, $Q$, for an SHB window of 1 year? 

\subsection{Proportions $=$ probabilities}  \label{SS:props=probs}
Sample-wise proportions of victims are just estimates of the probability of choosing a victim if one chooses a child at random from the population. The population is a semi-mythical concept, but we ignore that disagreeable fact. Probability is a more general concept then proportion. Regard annualized proportions $Q$ as population level probabilities, or estimates thereof.

\subsection{Probability Model}
Statistical models are expressed in terms of probability.
We develop a probability model for the study and use probabilities computed from the model as the $Q$'s in $FM_{3}$.
We need a flexible model that allows us to consider SHB in any SHB window. 

\subsection{Point Process}
Regard the SHBs as point events, i.e., having no duration. Ignore lethality, method, etc. 
This makes the events identical. Then a subject's history of SHB in a window can be summarized by the times, if any, at which he/she performed SHB during the time window. These event times, if any, are random. A theoretical gizmo that generates random times (possibly none) is called a ``point process''.
We model a subject's SHB history as a point process. 

A point process model gives us a broad framework we can use to compute all the means and standard deviations and probabilities we want concerning the SHB history.
With such a model we're not confined to any particular SHB window.
Now, we don't actually have the SHB history of any of the subjects.
So the point processes are ``latent''.
They constitute the back story of the data that we use to derive formulas that can be applied to the summary data presented in the paper.

\subsection{Poisson process}
\begin{figure}
      \epsfig{file = 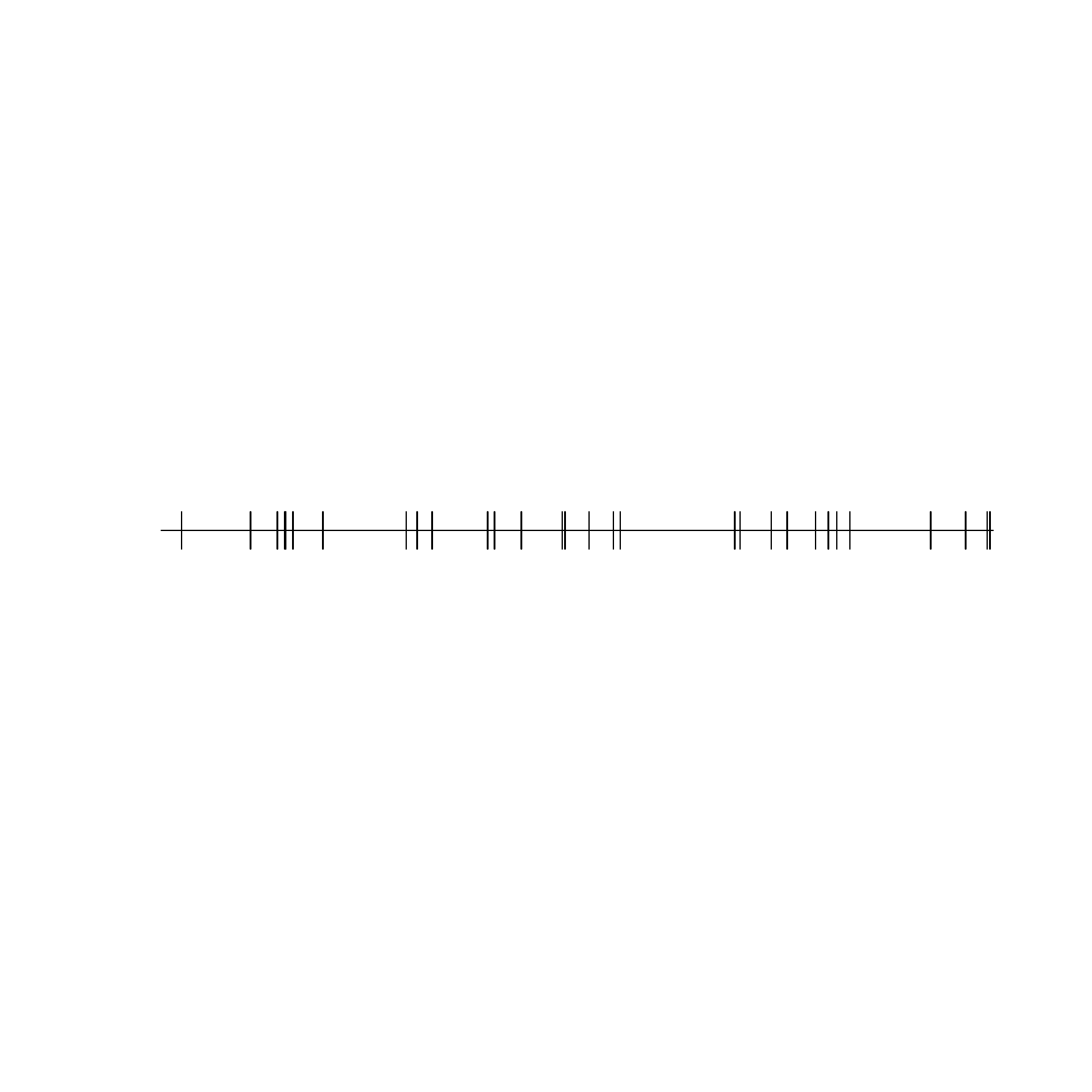, height = 1.2in, width = 6in, }  
      \caption{``Realization'' of a Poisson process on an interval.}    
             \label{F:PoissonProcess}
  \end{figure}

In line with Einstein's apocryphal advice we choose the simplest point process model: 
A stationary Poisson process. (`Stationary'' means the statistical properties of the process don't change with time. See figure \ref{F:PoissonProcess}.)
  
\subsection{Parametric Family}
\begin{figure}
      \epsfig{file = 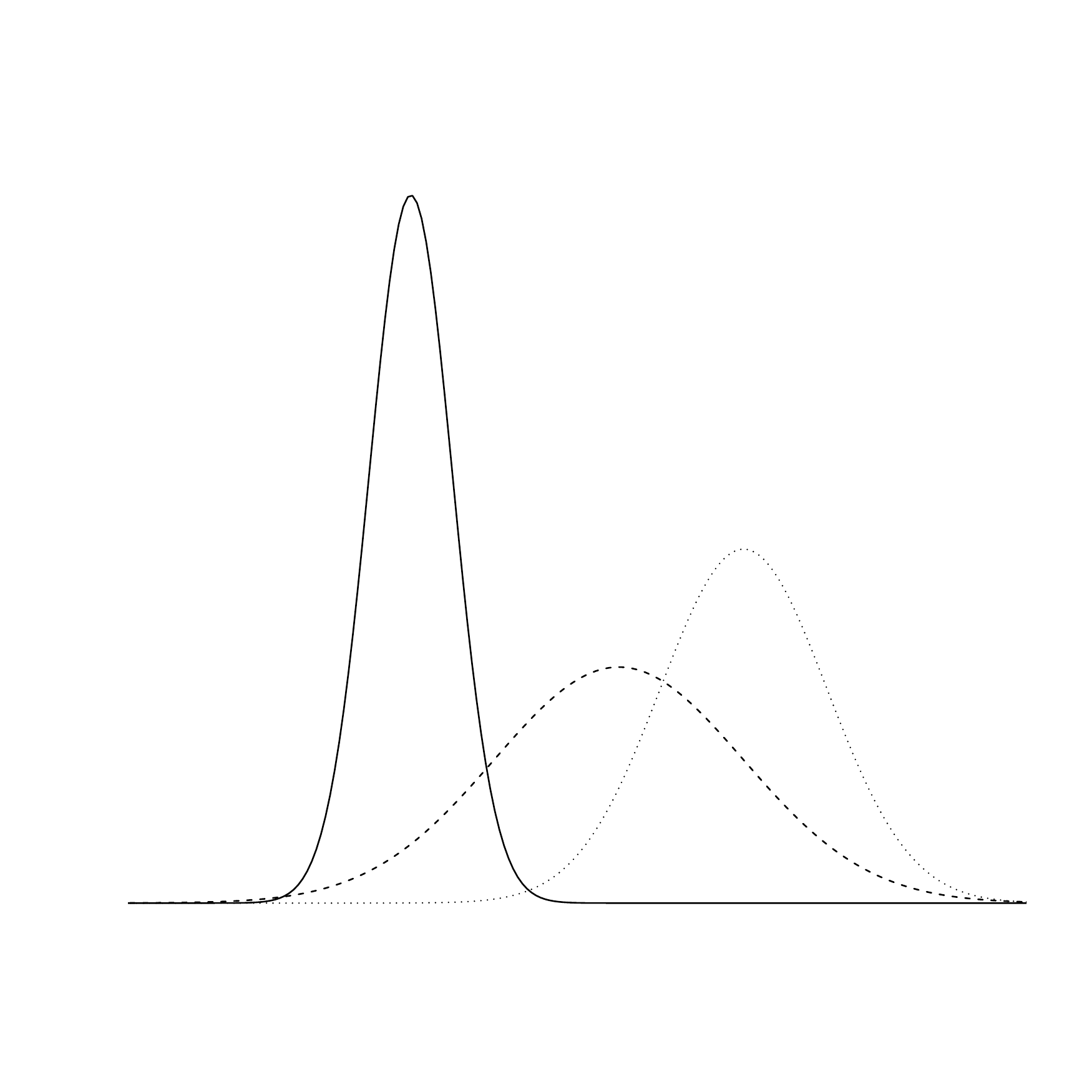, height = 4.4in, , }  
      \caption{Some normal curves.}    
             \label{F:normalCurves}
  \end{figure}

A common notion in statistics that we use incessantly is ''parametric family''. 
Example: Normal distribution. (See figure \ref{F:normalCurves}.)
The normal distribution isn't a single distribution.
It's a family of distributions each of which is specified by two numbers: The mean and standard deviation.

A family of distributions each of which is specified by a choice of some numbers is a ``parametric family''.
The numbers that specify a particular member of the family are called ``parameters''.
The normal distribution is a parametric family with parameters mean and SD.
Not every family of distributions is a parametric family.
(Unless you greatly extend the notion of ``parameter''.)

\subsection{Parametric and Nonparametric Inference}
In statistical inference one chooses some family, $\mcl{F}$, of probability distributions one member of which is supposed to capture the random nature of the data you have or expect to have.
You don't know which distribution in $\mcl{F}$ is the right one. 
You use the data to learn something about it.
If $\mcl{F}$ is a parametric family that's called ``parametric inference''
Examples: t-test and typical linear regression like ordinary least squares and mixed linear models.
If $\mcl{F}$ isn't a parametric family then you're doing ``nonparametric inference.''
Example: Wilcoxon or sign tests or Kaplan-Meier survival analysis.

\subsection{``Fitted model''}
Suppose you have some real world phenomenon that you want to model by a parametric family, say, the normal family. Problem: What are the parameter values?
Solution: You can't know them, but you can estimate them from data. 
Example: Sample mean and SD are estimates of the parameters of the normal distribution.
A probability distribution chosen from a parametric family by using empirical estimates as the parameter values is a ``fitted model''.

\subsection{Poisson Process Again}
The stationary Poisson process is a parametric family with one parameter: The ``rate''.
The rate is the average number of events the process puts in a unit of time.
If you know the rate of a Poisson process you can (in theory) compute probabilities, means, standard deviations, etc.\ for anything having to do with that process.

\subsection{Probability Model, First Version}
We model a subject's SHB history by three Poisson processes:
A pre-baseline process with (unknown) rate $\lambda$.
In follow-up, a process with (unknown) rate $\mu_{exp}$ for the experimental group and a third process with rate $\mu_{con}$ for the control group. In this model, the three Poisson processes are independent. We have to estimate the parameters from the data.

This model is a parametric family with three parameters: $\lambda$, $\mu_{exp}$, $\mu_{con}$.
Denote the model consisting of the three independent Poisson processes 
by $M_{1}(\lambda, \mu_{exp}, \mu_{con})$.
We will assume this model is true. Until we don't.

\subsection{Use Probability Model to Estimate $FM_{3}$}  \label{SS:use.prob.models}
With the fitted model we can estimate the annualized rates $Q_{exp}$ and $Q_{con}$ 
and therefore $FM_{3}$.
(But we have to figure out how to estimate the rates.)
Studies draw subjects from populations of children with very different rates of SHB.
I.e., the parameters $\lambda$, $\mu_{exp}$ , and $\mu_{con}$ differ among studies.

As a bonus, once we have computed $Q_{exp}$ and $Q_{con}$ we instantly get the relative risk: $Q_{exp}/Q_{con}$. Let's call this relative risk the "marginal relative risk''. But see section \ref{SS:NR.causal.model}.

\subsection{Doubts about $FM_{3}$}
Consider two studies, $A$ and $B$, in which a subject's chances of engaging in SHB post intervention is cut in half by the intervention. I.e., in both cases the relative risk is 1/2.
We might say that the two experimental interventions appear approximately equally effective.
But suppose study $A$ drew its subjects from a population that's twice as suicidal as that from which study $B$ drew its subjects. 
Then $FM_{3}$ for study $A$ will be double that for study $B$.
So by $FM_{3}$ standards, $A$ and $B$ are not equivalent.
So $FM_{3}$ can be incomparable between studies with different populations
The proportion scale is a poor scale for comparing the studies. 

\subsection{Cohen's and Relative risk.}
In meta-analysis people often use ``Cohen's d'' as the FM. Each subject experiences an intervention ``effect''. 
Defining what the ``effect'' is will require some care.
    \begin{equation}   \label{E:Cohen's.D.defn}
      \text{Cohen's d} =  \frac{\text{Avg of effect}}{\text{SD of effect}} .
    \end{equation}
Here, ``Average'' $=$ ``mean'' and ``SD'' means standard deviation (not standard error).

To interpret Cohen's d,  $FM_{3}$ is the starting point.
We will interpret $FM_{3} = Q_{flup,exp} - Q_{flup,con}$ as an average effect and $FM_{3}$ will become the numerator of Cohen's d. But $FM_{3}$ doesn't look like an average.

\subsection{Indicators}
Consider an event, $E$, say, that a child has engaged in SHB in the last year.
One can consider the proportion, $Q$, of children in the population for which this is true 
($=$ probability that this is true for a child randomly chosen from the population, subsection \ref{SS:props=probs}).
Consider the variable $1_{E}$ which is 1 for children with an attempt in the past year and 0 for all other children. $1_{E}$ is the ``indicator variable'' for the event $E$. So, yes, it's a variable. As such it has a mean.

Indicators are not uncommon: Whenever we code a dichotomous variable as 0 and 1 we're creating an indicator. Recall that $Q = $Probability of the event $E$. $1_{E}$ is the indicator of the event $E$. The punch line is:
    \begin{equation*}
      \text{Any probability is the mean of an indicator.} .
    \end{equation*} 
Specifically, $Q$ is the mean of $1_{E}$. Another fact about the mean: The difference of the means of two variables equals the mean of the difference of the variables..

For a given subject in the control group, let $1_{base}$ and $1_{flup,con}$ be the indicators of the the events the subject exhibits SHB during a one year period prior to enrollment and in the control condition in the follow-up period, respectlively. If the subject is in the experimental group $1_{flup,exp}$ is defined similarly to $1_{flup,con}$. We deduce that $1_{flup,exp} - 1_{flup,con}$ is -1, 0, or 1.
The difference, $1_{flup,exp} - 1_{flup,con}$, is the ``effect'' in the Cohen's d definition \eqref{E:Cohen's.D.defn}. And the average (mean) of this effect is nothing but
$Q_{flup,exp} - Q_{flup.con}$, i.e., $FM_{3}$.

\subsection{The sound of one hand clapping}
$1_{flup,exp}$ for a child $=$ 1 if he/she is randomized to the experimental group and then exhibits SHB in a one year period. 
$1_{flup,exp}$ $=$ 0 if he/she is randomized to the experimental group but doesn't exhibit SHB.
But what is $1_{flup,exp}$ if the child is not randomized to the experimental group?
We run into the same problem for $1_{flup,con}$.
So the effect, $1_{flup,exp} - 1_{flup,con}$ is not defined, in the sense we cannot acually determine its value for a given subject.

\subsection{How do clinicians think?}
To help find our way out of this quandry consider a clinician trying to decide between two mutually exclusive treatments A and B for a particular patient. (I.e., the patient can't get both.) How does the clinician ponder this question?
\begin{enumerate}
\item Scenario \ref{I:different.patients}:
``Hmm, suppose I prescribe treatment A for the patient and B for a comparable patient. Who would do better?'' \label{I:different.patients}
    \begin{itemize}
      \item Different treatments, different patients.
    \end{itemize}
\item Scenario \ref{I:same.patient}:
``Hmm, suppose I prescribe treatment A for the patient. How would he/she do in comparison to how he/she would do if I prescribed treatment B for him/her?'' \label{I:same.patient}
    \begin{itemize}
      \item Different treatments, same patient.
    \end{itemize}
\end{enumerate}
Scenario \ref{I:same.patient} describes the clinician's thinking. Unfortunately, that scenario is physically impossible. It can only exist as a thought experiment.

\subsection{Neyman-Rubin causal model for control studies} \label{SS:NR.causal.model}
We want to formalize the clinician thought experiment \ref{I:same.patient}.
The idea is we imagine giving the experimental intervention to the subject and at the same time in an alternate universe give the control intervention to the \emph{same} subject, then ask what would be the difference in the two outcomes.
In this fantasy we can nail down causality because confounds are perfectly adjusted for.
Why bother thinking about this physically impossible scenario?
Because it makes clear what we'd really like to know so that we can apply the right formulas to the actual data.
What we really want to know is how the child would respond to the experimental intervention compared to how the same child would respond to the control intervention.

This is a standard viewpoint in the analysis of clinical trials (Neyman \cite{jN1923.CausalModel}, Rubin \cite{dR2006.CausalEffectsBook}) and we adopt it. In this fantasy $1_{flup,con}$ and  $1_{flup,exp}$ are both always defined because the child is always in the both arms. Now the effect for an individual child in the study is well-defined:
Effect $=$ $1_{flup,exp}$ - $1_{flup,con}$.
Effect is a random variable taking the values -1, 0, or 1.
So now we can interpret $FM_{3}$ as the mean of an effect:
$FM_{3}$ $=$ Avg of ($1_{flup,exp}$ - $1_{flup,con}$)

Recall 
    \begin{equation*}
      \text{Cohen's d} = \frac{\text{Mean of Effect}}{\text{SD of Effect}}.
    \end{equation*}
We know what effect is and how to compute its mean:
Mean Effect $=$ $FM_{3}$ $= Q_{flup,exp} - Q_{flup,con}$. 
Now we have to compute the SD of Effect . Recall 
    \begin{multline*}
      \text{Effect } = 1_{flup,exp} - 1_{flup,con} \;  \text{ so } \\
    ( \text{SD of Effect } )^{2} = ( \text{SD of } 1_{flup,exp} )^{2} + ( \text{SD of } 1_{flup,con})^{2}  \\
    - 2 ( \text{SD of } 1_{flup,exp} ) \times ( \text{SD of } 1_{flup,con}) 
                         \times (\text{correlation of }  1_{flup,exp} \text{ and } 1_{flup,con} ).
    \end{multline*}

Correlation of $1_{flup,exp}$ and $1_{flup,con}$ is the correlation between the $1_{flup}$'s in different fantasy universes. Will there be a correlation? Yes, because it's the same child in both universes.
This correlation is due to variability among children.

Relative risk can also be computed in this spirit. For each subject we can compute the annualized probability of SHB for both their experimental and control versions. Call the the quotient the ``Neyman-Rubin relative risk''. 

\subsection{Correlation} \label{SS:correlation}
So the correlation of $1_{flup,exp}$ and $1_{flup,con}$ may not be 0, but how do you even begin to assign a value to it? We start with an easier problem, what's $r$, the correlation between $1_{base}$ and $1_{flup,con}$?
$1_{base}$ $=$ 1 if the child had any SHB's in the 1 year prior to baseline. Otherwise, 0.
$1_{flup,con}$ $=$ 1 if the child had any SHB's in 1 year of follow-up after (during) getting the control intervention. Otherwise, 0.

Might $r$ be zero? Tendency toward SHB obviously varies from child to child at baseline.
But in follow-up, too. SHB is trait-like so $r=0$ seems unlikely. In fact, it is almost certain that $r > 0$.

Even if SHB is reduced by the control intervention in every subject, we posit that a child with higher than average SHB risk at baseline will tend to also have higher than average risk in follow-up.
Similarly for children with lower than average risk at baseline.
This shared variability pre- and post- will induce a positive correlation between $1_{base}$ and $1_{flup,con}$. So what is $r$?
The papers we're using for our meta-analysis don't say anything about $1_{base}$ 
and $1_{flup,con}$. In fact, with one exception the papers don't say anything about pre- post- dependence. Editorial comment: When you report the results of an intervention pre- and post- don't just give the baseline and follow-up means and SD's. Please also give the correlation!

The paper doesn't say anything about $r$ but $r$ has to be something.
Assuming $r=$ 0 is not agnostic. To make up for this omission we hypothesize values of $r$:
$r_{hyp}=$ 0.1, 0.3, or 0.5
We do the calculations assuming each of those values for $r_{hyp}$.

Turns out our current model cannot accommodate correlation.
Recall that our current model consists of one Poisson process for baseline, follow-up in the control group, follow-up in the experimental group, all three independent of each other. So in this model $r = 0$.

We can introduce non-zero correlation by modulating Poisson rates. Let $R$ be a positive variable that measures the SHB trait in a subject.
$R > 1$ means SHB tendency higher than average. 
$R = 1$ means average SHB tendency.
$R < 1$ means SHB tendency lower than average.
$R$ varies from subject to subject.
There are three rates of SHB in the Poisson model: $\lambda$, $\mu_{con}$, $\mu_{exp}$.
Replace them by $R \lambda$, $R \mu_{con}$, $R \mu_{exp}$.
So $R$ modulates the propensity for SHB on a subject-wise basis.

So now instead of three Poisson processes we have three ``mixed Poisson processes'' or ``Cox processes.''
The rates vary randomly across subjects but we may assume that the rates averaged over subjects are still $\lambda$, $\mu_{con}$, $\mu_{exp}$.
Those remain the population values.

Recall that $1_{base}$ and $1_{flup,con}$ both belong to the same subject.
They share variability because they share the same $R$, which varies across subjects.
This gives rise to correlation between them. Thus, our new Cox model allows for correlation.

\begin{figure}
      \epsfig{file = 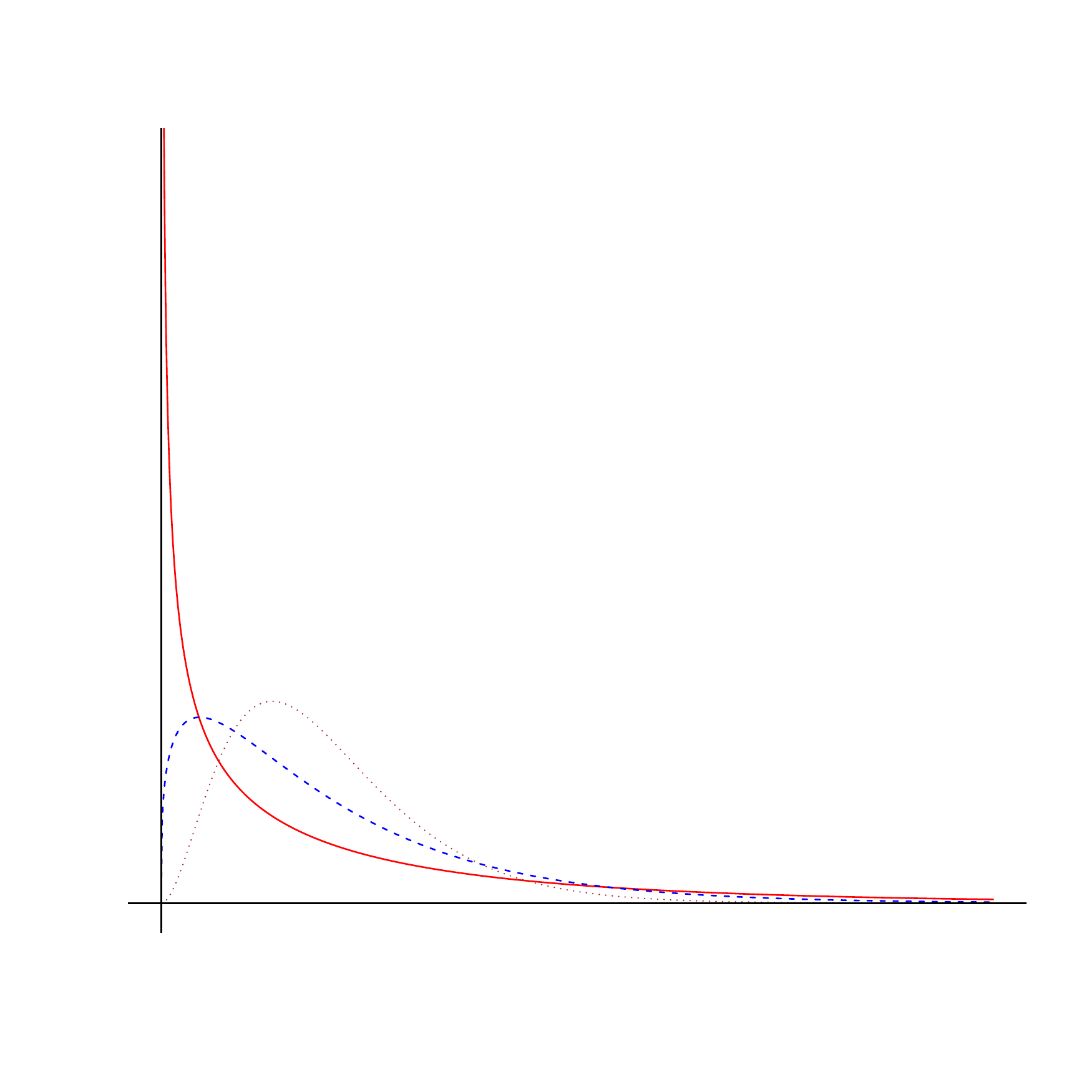, height = 4.4in, , }  
      \caption{Some gamma curves.}    
             \label{F:gammaCurves}
  \end{figure}

$R$ is random. We don't know its distribution, but we can pick a parametric family of distributions to which we may hypothesize the distribution belongs.
We pick the family of ``gamma distributions''.
Like the normal family, the gamma family of distributions is a 2 parameter family.
The parameters are sometimes called ``shape'' and ``scale''.
You may have used the gamma family without knowing it:
The chi-squared distribution is a one parameter subfamily of the gamma.
The parameter for chi-squared is called ``degrees of freedom''.
In our Cox model there is a redundancy among the parameters so that we only need to consider gamma distributions with parameters $\alpha$, $1/\alpha$, where $\alpha$ is any positive number.
Thus, one number suffices to determine the distribution.
So we're working with a one-parameter gamma subfamily. Figure \ref{F:gammaCurves} shows some density curves in this family.

Our new model consists of three Cox processes, with random rates $R \lambda$, $R \mu_{con}$, $R \mu_{exp}$, where $R$ comes has a gamma distribution with parameters 
$(\alpha, 1/\alpha)$. 
Denote that model by
$M_{2}(\lambda, \mu_{con}, \mu_{con}, \alpha)$ .
A choice of numerical values for the four parameters  determines a distribution for any random quantity associated with the model.
E.g., any positive values for $\lambda$, $\mu_{con}$, $\mu_{con}$, and $\alpha$ determine the correlation between $1_{base}$ and $1_{flup,con}$ .

$M_{2}(\lambda, \mu_{con}, \mu_{con}, \alpha)$ describes a toy world in which there are only 
instantaneous identical events occurring in time
under three different conditions: Baseline, follow-up control arm, follow-up experimental arm.

Remember correlation of $1_{flup,exp}$ and $1_{flup,con}$?
We needed it in order to calculate the SD of correlation of 
$1_{flup,exp}$ -$1_{flup,con}$ which is to be the denominator in Cohen's d.
$1_{flup,exp}$ and $1_{flup,con}$ share variability because they pertain to the same subject and hence share the same $R$. This gives rise to correlation between $1_{flup,exp}$ and $1_{flup,con}$ .
Within our Cox model it is easy to calculate that correlation.
Given values of $\lambda$, $\mu_{con}$, $\mu_{exp}$, and  $\alpha$, we can compute Cohen's d for that member of our Cox family of distributions.
But we don't know $\lambda$, $\mu_{con}$, $\mu_{exp}$, and  $\alpha$.
Use data to estimate them.
This is the statistical inference problem: Use data to learn about unknown parameters.

We have 4 unknown parameters: $\lambda$, $\mu_{con}$, $\mu_{exp}$, and  $\alpha$.
To estimate them we need at least 4 pieces of information: Recall that $q_{base}$ is the proportion of subjects who exhibited SHB during the baseline SHB window. (See? I promised you I would use $q_{base}$.) SHB window is an interval of time (of possibly random length, subsection \ref{SS:lifetime.SHB} and subsubsection \ref{SSS:truncated.normal}).
$q_{flup,con}$ and $q_{flup,exp}$ are the proportions of subjects who exhibited SHB during the specified SHB window at follow-up in the control arm and experimental arm, respectively.
The fourth piece of information is $r_{hyp} =$ 0.1, 0.3, or 0.5

For every choice of $\lambda$, $\mu_{con}$, $\mu_{exp}$, and  $\alpha$, and also using the lengths of the SHB windows, we can compute the corresponding theoretical population values of $q_{base}$, $q_{flup,con}$, $q_{flup,exp}$, and $r$
Call them $q_{base}(\lambda, \mu_{con}, \mu_{exp}, \alpha)$, 
$q_{flup,con}(\lambda, \mu_{con}, \mu_{exp}, \alpha)$, 
$q_{flup,exp}(\lambda, \mu_{con}, \mu_{exp}, \alpha)$, and 
$r(\lambda, \mu_{con}, \mu_{exp}, \alpha)$. Please distinguish these theoretical $q$s and $r$ from the observed values of $q_{base}$, $q_{flup,con}$, $q_{flup,exp}$, reported in the paper and the hypothesized value $r_{hyp}$.

$q_{base}(\lambda, \mu_{con}, \mu_{exp}, \alpha)$, etc., are given by formulas involving 
$\lambda$, $\mu_{con}$, $\mu_{exp}$, and  $\alpha$ and the relevant SHB windows.
Since we're using a point process model, we can derive formulas for any SHB window. Example: Suppose the length of the baseline SHB window is $T$. Then it turns out that
    \begin{equation}  \label{E:q.base.formla}
        q_{base}(\lambda, \mu_{con}, \mu_{exp}, \alpha) 
          = 1 - \left( \frac{\alpha}{\alpha + \lambda T} \right)^{\alpha} .
    \end{equation}

Using the $M_{2}(\lambda, \mu_{con}, \mu_{con}, \alpha)$ model we can also estimate the Neyman-Rubin relative risk defined in section \ref{SS:NR.causal.model}. For a given number $s > 0$ one can compute the relative risk corresponding to $R = s$. Call that $RR(s)$. Now replace $s$ by the random variable $R$ and average over $R$.

\subsection{Lifetime SHB} \label{SS:lifetime.SHB}
In many papers $q_{base}$ pertains to lifetime SHB. What's the length of the lifetime SHB window?
Not literally lifetime: We defined SHB ``lifetime'' as begining at age 10. A child's lifetime SHB window ends at baseline. So lifetime SHB window length $=$ (age at baseline) - 10. But children are different ages at baseline so length of the lifetime SHB window varies from child to child. 

So to estimate the value of, e.g., $q_{base}(\lambda, \mu_{con}, \mu_{exp}, \alpha)$ for the whole sample we have to average over $T$ in equation \eqref{E:q.base.formla}. Essentially, average over age at baseline.

How do he compute this average? We can't average the right hand side of equation \eqref{E:q.base.formla} for a study by plugging in the actual ages of the subjects at baseline and averaging: We don't have subject level data. But do have age range, mean age, and age SD.
``Range'' means minimum and maximum baseline ages for children in the study.
Use that and the age mean and SD to fit a model age distribution based on normal distribution (subsubsection \ref{SSS:truncated.normal}). We use that fitted distribution to compute average (mean). This gives a more complicated formula for 
$q_{base}(\lambda, \mu_{con}, \mu_{exp}, \alpha)$. Anyway, hopefully this gives you some feeling for how we derive formulas for 
$q_{base}(\lambda, \mu_{con}, \mu_{exp}, \alpha)$, 
$q_{flup,con}(\lambda, \mu_{con}, \mu_{exp}, \alpha)$,
$q_{flup,exp}(\lambda, \mu_{con}, \mu_{exp}, \alpha)$, and
$r(\lambda, \mu_{con}, \mu_{exp}, \alpha)$. For the details, see section \ref{S:the.math}.

So now we have four equations in four unknowns:
    \begin{align}  \label{E:4.eqns.in.4.unkns}
        q_{base}(\lambda, \mu_{con}, \mu_{exp}, \alpha)  &=  q_{base}, \notag \\
        q_{flup,con}(\lambda, \mu_{con}, \mu_{exp}, \alpha) &= q_{flup,con}, \\
        q_{flup,exp}(\lambda, \mu_{con}, \mu_{exp}, \alpha) &=q_{flup,exp}, \text{ and } \notag \\
        r(\lambda, \mu_{con}, \mu_{exp}, \alpha) &= r_{hyp} , \notag 
    \end{align}
where the $q_{base}$, $q_{flup,con}$, $q_{flup,exp}$ on right hand sides of equations are observed values from the paper and $r_{hyp}$ is the hypothesized correlation between baseline and control follow-up, 0.1, 0.3, or 0.5.

We find values of $\lambda$,$\mu_{con}$,$\mu_{exp}$, and $\alpha$, for which the equations are true or at least close to being true. These are our ``estimated parameter values''. We indicate estimates by putting a ``hat'' ``$\hat{\;}$'' over the symbol, like this: $\hat{\alpha}$.
The estimated parameters correspond to one member of the parametric model family.
 The corresponding model is the ``fitted model''.
Use fitted model to compute Cohen's d.

We actually do this! Here's a snippet from the output in analyzing Asarnow \emph{et al} \cite{jrAetAl2017.CBfamTSuiPrevent}. 
Outcome is attempt. Our parameter estimates are $\hat{\alpha} =  0.911$, $\hat{\lambda} = 0.527$, 
  $\hat{\mu}_{C} = 0.908$ , $\hat{\mu}_{E} = 1.161 \times 10^{-10}$. 
  Here's how close these estimates come to solving the four equations (with $r_{hyp} = 0.3$):
  $1.65 \times 10^{-24}$. Pretty darn good. From the fitted model we can compute:
     
      \begin{table}[h]
	\begin{tabular}{ | p{.5in} | p{.5in} | p{.5in} | p{.5in} |  p{.5in} | p{.5in} | p{.5in} | p{.5in} |  }
	\hline
           $Q_{base}$ &  $Q_{flup,con}$    &  $Q_{flup,exp}$      &   $r$ & CohenD & SD    &
             $M \, RR$    &  $NR \, RR$ \\ \hline
           0.340  & 0.465 & 0.093 & 0.300 & -0.695 & 0.535 & 0.2  & 0.283 \\ \hline
	\end{tabular}
     \end{table}

Here, ``$Q_{base}$'' is the estimated annualized baseline probability, i.e., the estimated probability that a subject from the population from which the sample was drawn would make an attempt in the year prior to enrollment in the study. (These numbers are estimates, but we haven't bothered putting hats on them. By our definition \eqref{E:Cohen's.D.defn} the Cohen's d of a successful trial will be negative. In the paper we changed the sign to make it conform to the expectations of the typical reader: Positive values are good.) ``$M \, RR$'' is the "marginal relative risk'' defined in section \ref{SS:use.prob.models} and ``$NR \, RR$'' is the ``Neyman-Rubin relative risk'' defined in section \ref{SS:NR.causal.model} and discussed further in section \ref{SS:correlation}.
  
\subsection{Inclusion, exclusion} \label{SS:inc-ex}
$Q_{base} = 0.340$ seems kind of high for a baseline attempt probability. What gives?
Asarnow's inclusion criteria:
``Inclusion criteria required: a recent (past 3 months) SA (suicide attempt); or NSSI (nonsuicidal self-injury) as primary problem, with the additional requirement of repetitive SH (self-harm), defined as $\geq$ 3 lifetime SH episodes.''

Asarnow's study sample was intentionally biased toward children at high risk. The algorithm took the sample as representative of some population. And the sample was representative, but of a somewhat skewed population. Such biased samples can be thought of in two ways. 
Inclusion-exclusion extracts biased samples from a ``true'' population or inclusion-exclusion extracts representative samples from a skewed population.

This wouldn't be a big problem if all the studies used the same inc-ex criteria. Unfortunately, the criteria vary wildly from study to study. 
E.g., Asarnow recruited high risk children, but for other studies history of attempt is an exclusion. 
Varying inclusion-exclusion criteria is a completely artifactual difference among the studies.

On a practical level, our model $M_{2}$ naively applied can break down in the presence of some inc-ex criteria. Is there some statistical way we can compensate for the diversity of inclusion-exclusion criteria? Sometimes, yes, by not applying $M_{2}$ naively.

\subsubsection{``Clean'' criteria} \label{SSS:cleanliness}
The inclusion-exclusion criteria involve lots of psycho-social dimensions: Varieties of SHB, ideation, family situation, IQ, present police custody. 
If we had subject-level data we could try to model some of these other measures. But we don't.
There's no place for all these dimensions in the simple world of our statistical model $M_{2}$.

Sometimes we can adjust within the confines of $M_{2}$. Say that an inclusion or exclusion criterion is ``clean'' if it only involves the SHB we've chosen as our outcome.
Example: Summary data on attempt is available in a paper and the study excluded children with a history of attempt.
Example: We have to settle on NSSI as our outcome and presence of NSSI in the last three months is an inclusion criterion. By contrast look again at Asarnow's criteria: `Inclusion criteria required: a recent (past 3 months) SA (suicide attempt); \emph{OR} NSSI (nonsuicidal self-injury) as primary problem, with the additional requirement of repetitive SH (self-harm), defined as $\geq$ 3 lifetime SH episodes.'' (Emphasis added.) Knowing a child's history of SHB is not enough to tell us whether they would be included or excluded from the study. Those criteria are not clean.

A clean criterion can be described in terms of the simple world of our model. The way we do this is to model the clean criterion in the model world. If all studies had the same clean criteria life would be simpler. But even clean criteria differ among studies.
To make the Cohen's d's for those studies comparable we compute, within the model world, what the Cohen's d's would be with the clean criteria removed, i.e., if the investigators had dropped that criterion. This makes the Cohen's d's more comparable across studies. 

We can compute probabilities, means, SD, etc. for anything that can happen in the model world.
In particular, for anything taking place conditional on clean criteria being satisfied. 
In effect, we adjust, we remove the clean criteria.
This leads to modified ``4 equations in 4 unknowns''.

Modeling inclusion/exclusion criteria was a huge pain. Most of section \ref{S:the.math} is taken up with deriving formulas that model inclusion/exclusion in various studies.

An example where we can adjust is the ``Youth Aware'' portion of Wasserman \emph{et al} \cite{dWcwHcWmW2015.SchoolBasedSuicidePrevention}. Here the outcome is attempt and
history of attempt was an exclusion. That's not the only inc-ex criterion. To quote the paper, ``All pupils who reported suicide attempts ever, or severe suicidal ideation in the past 2 weeks before the baseline assessment, and those with missing data regarding these two variables were not included in the final analysis.'' In the $M_{2}$ model we cannot model ideation or missingness, but we know that if a subject had a history of attempt they were excluded. 

Luckily, Wasserman and co-authors were kind enough to provide information about the subjects who were excluded because of attempt history. We used those numbers to measure how suicidal was the population from which they were sampling. In the $M_{2}$ model one can derive new formulas -- the formulas are derived in subsubsection \ref{SSS:zero.baseline} below -- 
for $q_{base}(\lambda, \mu_{con}, \mu_{exp}, \alpha)$, 
$q_{flup,con}(\lambda, \mu_{con}, \mu_{exp}, \alpha)$,
$q_{flup,exp}(\lambda, \mu_{con}, \mu_{exp}, \alpha)$, and
$r(\lambda, \mu_{con}, \mu_{exp}, \alpha)$ and after solving the resulting four equations 
(with $r_{hyp} = 0.3$) we get the estimates: 
$\hat{\alpha}= 0.043$, $\hat{\lambda}= 0.012$, $\hat{\mu}_{con} = 0.099$, 
$\hat{\mu}_{exp}= 0.074$. Using the fitted model we compute further estimates;

      \begin{table}[h]
	\begin{tabular}{ | p{.5in} | p{.5in} | p{.5in} | p{.5in} |  p{.5in} | p{.5in} | p{.5in} | p{.5in} |  }
	\hline
           $Q_{base}$ &  $Q_{flup,con}$    &  $Q_{flup,exp}$      &   $r$ & CohenD & SD    &
             $M \, RR$    &  $NR \, RR$ \\ \hline
           0.010  & 0.050 & 0.042 & 0.300 & -0.039 & 0.204 & 0.841  & NA \\ \hline
	\end{tabular}
     \end{table}

(We were unable to compute $NR \, RR$ for this study. Wasserman's wasn't the only study in which this occurred.) So even though no subjects with a history of attempt were enrolled, we still get a non-zero value of $Q_{base}$. That's because our formulas take into account the exclusion criterion.

For two papers we had to perform this kind of adjustment once on the ideation side. See section \ref{SSS:Diamond.ideation}.

\subsection{SE}
We also want a SE for Cohen's d and the relative risks. We use them to weight papers in regression analysis at the end. We computed SE's by means of a ``parametric bootstrap''. (This is where sample size enters the analysis.) We used the fitted model to generate 1,000 artificial data sets.   For each artificial data set we then, in a manner identical to what we did in the analysis of the actual meta-data, computed Cohen's d and the relative risks. The result is 1,000 Cohen's d's and 1,000 relative risks, of each type. We computed the SD of each. Those are the SEs. Thus, the SDs of the sets of the 1,000 statistics are the SE's of the estimates we got from the actual data.

\emph{This concludes the description of the analysis of behavior meta-data for one paper.} (See section \ref{S:ideation} for discussion of how the ideation meta-data were analyzed.)

\section{Some Math} \label{S:the.math}

\subsection{Introduction} \label{S:intro}
This section describes some theory underlying the methods used by the author for the Itzhaky \emph{et al}  \cite{lIetAl2021.SuicidePreventionMetaAnalysis} meta-analysis. Section \ref{S:basic.concepts} started life as a Power Point presentation. This section was created to record derivations of formulae used in the R programs that were employed to do the calculations. It grew gradually and the result was somewhat ungainly. It's been revised, but it still needs work. Hopefully, in its present form it's nonetheless readable.

We analyze behavior and ideation separately because they are measured very differently. Behavior takes the form of discrete events while ideation varies continuously over time. However, we made the meta-analytic output for the two sides of SHB comparable, viz., as a Cohen's d. (Relative risk doesn't make sense for a quantitative measure like ideation.) This makes it possible to plot behavior and ideation together, if we wanted to. (But, alas, we didn't.) 

\subsection{Behavior} \label{S:behavior}
The analysis of behavior is far trickier than that of ideation. Much of the complexity is due with correcting for inclusion-exclusion criteria as described in section \ref{SS:inc-ex}. The papers exhibit a wide variety of ``clean'' criteria (section \ref{SSS:cleanliness}). For each such criterion, appropriate formulas, as in sections \ref{SS:correlation}, \ref{SS:lifetime.SHB}, and \ref{SSS:cleanliness}, have to be derived. Most of the work described here has to do with those derivations.

\subsubsection{Gamma mixture of Poissons} \label{SSS:gamma.mixture}
Here we derive generic formulas for the Cox process model $M_{2}$ described in section \ref{SS:correlation}. 

$R$ is chosen independently for each subject. Thus, $R$ represents the subject specific contribution of suicidality risk. Assume $R$ follows a $\Gamma(\alpha, \beta)$ distribution (Hoel, Port, and Stone \cite[p.\ 129]{pgHscPcjS71.IntroToProbThy}) 
for some $\alpha, \beta > 0$. Thus, the SHB counts in any time period follows a  negative binomial distribution.

Let $\lambda_{1}$, $\lambda_{2}$ be the population wide Poisson rates for the pre- and post-intervention event counts (per unit time), respectively, for one of the study arms. Then for a given subject the conditional (on $R$) pre- and post-intervention event count (per unit time) rates will be $R \lambda_{1}$, $R \lambda_{2}$, respectively. Now, $\lambda_{1}$ is the average pre- rate across the population. That means 
    \begin{equation} \label{E:ER.=.1}
      E R = 1 .
    \end{equation}. 
Hence, by Hoel, Port, and Stone \cite[Equation (2), p.\ 177]{pgHscPcjS71.IntroToProbThy} we have $R \sim \Gamma(\alpha, \alpha)$ 
for some $\alpha > 0$. On the same page of \cite{pgHscPcjS71.IntroToProbThy}, we find that consequently,
    \begin{equation*}
        Var \, R = 1/\alpha.
    \end{equation*}
Thus, small $\alpha$ corresponds to high heterogeneity among subjects.

Let $N_{1}, N_{2}$ be the pre- and post-intervention event counts over time intervals 
$T_{1}, T_{2} > 0$. Then, given $R$, $N_{i}$ is Poisson with rate $R \lambda_{i} T_{i}$. Since $N_{1}, N_{2}$ are based on non-overlapping time periods, conditional on $R$, they are independent. Therefore (\cite[Example 9, p.\ 56]{pgHscPcjS71.IntroToProbThy}), the unconditional probability that $N_{i} > 0$ is given by
    \begin{equation*}
        q_{i} := q_{i}(\alpha) := Prob \{ N_{i} > 0 \} = 1 - Prob \{ N_{i} = 0 \} 
          = 1 - E \exp \{ - R \lambda_{i} T_{i} \}.
    \end{equation*}
Using the formula for the moment generating function of a gamma random variable (\cite[Equation (2), p.\ 198]{pgHscPcjS71.IntroToProbThy}) we then find
    \begin{equation}  \label{E:qi.formla}
        q_{i} = 1 - E \exp \{ - R \lambda_{i} T_{i} \}
          = 1 - \left( \frac{\alpha}{\alpha + \lambda_{i} T_{i}} \right)^{\alpha} .
    \end{equation}
(This is just a reprise of \eqref{E:q.base.formla}. See \eqref{E:unconditional.prob.of.0}.) Note that for $\alpha, T_{i} > 0$ given, allowing $\lambda_{i} $ to vary between 0 and $\infty$, we can make $q_{i}$ any number in $(0,1)$. 

Examine the behavior of $q_{i}(\alpha)$ as $\alpha \uparrow \infty$.
    \begin{equation*}
        q_{i}(\alpha) = 1 - \left( \frac{1}{1 + \lambda_{i} T_{i}/\alpha} \right)^{\alpha} 
          = 1 - \frac{1}{ \left( 1 + \lambda_{i} T_{i}/\alpha \right)^{\alpha} } 
            = 1 - \left[ \frac{1}{ \left( 1 + \lambda_{i} T_{i}/\alpha \right)^{\frac{\alpha}{\lambda_{i} T_{i}}} } \right]^{\lambda_{i} T_{i}}.
    \end{equation*}
By Rudin \cite[Theorem 3.31, p.\ 55]{wR64.PMA}, we see that as $\alpha \uparrow \infty$ this last expression converges to $1 - \exp \{ - \lambda_{i} T_{i} \}$, the probability that a Poisson r.v.\ with mean $\lambda_{i} T_{i}$ is positive. This is consistent with homogeneity of subjects.

Using the theorem in Feller \cite[p.\ 89]{wF57.feller1},
    \begin{multline}  \label{E:Prob.X1=0.X2=0}
        Prob \{ N_{1} = 0 \text{ or } N_{2} = 0 \} \\
        \begin{aligned}
	  &= Prob \{ N_{1} = 0 \} + Prob \{ N_{2} = 0 \} - Prob \{ N_{1} = 0, N_{2} = 0 \} \\
          &= Prob \{ N_{1} = 0 \} + Prob \{ N_{2} = 0 \} - Prob \{ N_{1} + N_{2} = 0 \} .
        \end{aligned}
    \end{multline}

Let $p_{12} := p_{12}(\alpha) := Prob \{ N_{1} + N_{2} > 0 \}$, we have by \eqref{E:qi.formla},
    \begin{equation*}
        p_{12} = 1 - \left( \frac{\alpha}{\alpha + \lambda_{1} T_{1}  + \lambda_{2} T_{2}} \right)^{\alpha}.
    \end{equation*}
Define $q_{12} := q_{12}(\alpha) := Prob \{ N_{1} > 0 \text{ and } N_{2} > 0 \}$. Then, by \eqref{E:Prob.X1=0.X2=0}, 
    \begin{multline*}
    	 q_{12} = 1 - Prob \{ N_{1} = 0 \text{ or } N_{2} = 0 \} \\
	  = \bigl( 1 - Prob \{ N_{1} = 0 \} \bigr) + \bigl( 1 - Prob \{ N_{2} = 0 \} \bigr) 
	    - \bigl( 1 - Prob \{ N_{1} + N_{2} = 0 \} \bigr) \\
                 = q_{1} + q_{2} - p_{12}.
    \end{multline*}

Define variables $S_{i} := 1_{ \{ N_{i} >0 \} }$ ($i = 1,2$). Then, $E S_{i} = q_{i}$ and 
$E (S_{1} S_{2}) = q_{12}$. Thus,
    \begin{equation*}
        q_{12} = E (S_{1} S_{2}) \leq \min \{ q_{1}, q_{2} \} .
    \end{equation*}

    \begin{equation*}
        Var \, S_{i} = q_{i} - q_{i}^{2} = q_{i} (1 - q_{i}) \text{ and } Cov ( S_{1} S_{2} ) 
          = q_{12} - q_{1} q_{2}. 
    \end{equation*}
Hence, the correlation between $S_{1}$ and $S_{2}$ is given by
    \begin{equation}   \label{E:gamma.mixture.correlation}
        Corr (S_{1}, S_{2}) =  \frac{q_{12} - q_{1} q_{2}}{ \sqrt{ q_{1} (1 - q_{1}) q_{2} (1 - q_{2}) } } 
          \leq \frac{ \min \{ q_{1}, q_{2} \}  - q_{1} q_{2} }{ \sqrt{ q_{1} (1 - q_{1}) q_{2} (1 - q_{2}) } }
    \end{equation}
This is the same as the ``Phi coefficient of association'' or "Matthews correlation coefficient'' (Wikipedia). 

We have performed calculations for $N_{1}, N_{2}$ the pre- and post-intervention event counts over time intervals of lengths $T_{1}, T_{2} > 0$. Now suppose $N_{2C}, N_{2T}$ are post-intervention event counts over time intervals $T_{2C}, T_{2T} > 0$ for hypothetical experiments in which the patient independently and counterfactually receives \emph{both} the control and treatment interventions. 
Then for given population values $\lambda_{2C}$ and $\lambda_{2T}$ and conditional on $R$, the variables $N_{2C}, N_{2T}$ are independent Poisson r.v.'s 
with conditional means $R \lambda_{2C} T_{2C}$ and $R \lambda_{2T} T_{2T}$, respectively. Then the formulas derived above translate immediately to this new situation.

\subsubsection{Truncated Gaussian distribution}  \label{SSS:truncated.normal}
$T_{1}$ is the amount of pre-enrollment time over which the self-harm is recorded. Sometimes the investigators specify a specific time period, a year, say. But often $T_{1}$ means "lifetime". Operationally, that is the age of the subject at enrollment minus the age at which the subject becomes at risk. We take that age to be 10. But usually subjects entering the study have ages anywhere in a given range, e.g., 12 to 17. Therefore, $T_{1}$ will be random. We discuss the choice of distribution for $T_{1}$ here. 

A principled choice of the distribution of $T_{1}$ can be obtained by using the principle of ``maximum entropy''. (See Wikipedia article, "Principle of maximum entropy".) Suppose the age range of subjects is constrained to lie in an interval $(10+\sigma, 10+\sigma+\tau)$, where $\sigma, \tau > 0$.

The typical paper gives proportions of subjects in each group who have performed self-injurious acts during the baseline period and the treatment period. And the baseline period is lifetime. Now we regard the age of risk for adolescents as beginning at age 10. So the baseline period extends from age 10 to the age at which the subject is enrolled in the study. This varies from subject to subject. The mean and standard deviations of the ages are given in the paper. 

We seek the density, $f$, that maximizes entropy subject to the restraints, first, that it is in fact a density, i.e., integrates to 1. Second, $f$ has the given mean and SD. And third that its support lies in the interval $(0, \tau)$. The entropy functional is
    \begin{equation*}
        H(f) := - \int_{0}^{\tau} f(x) \log f(x) \, dx.
    \end{equation*}
If $f$ is a maximizer then if we perturb $f$ in the direction of a function $\rho$ then $H$ can only be reduced. Thus,
    \begin{equation}  \label{E:.partial.H.w.r.t.t.= 0}
        \frac{\partial H(f + t \rho)}{\partial t} \vert_{t = 0} = 0.
    \end{equation}

But not all directions are legal. Let $\mu$ and $\sigma$ be the target mean and SD then 
    \begin{equation}  \label{E:moments.of.f}
        \int_{0}^{\tau} f(x) \, dx = 1, \; \int_{0}^{\tau} x f(x) \, dx 
          = \mu; \; \int_{0}^{\tau} (x - \mu)^{2} f(x) \, dx = \sigma^{2}.
    \end{equation}
Thus, in order for $f + t \rho$ to satisfy the constraints we must have 
    \begin{multline} \label{E:density.constraints}
        \int_{0}^{\tau} \bigl[ f(x) + t \rho(x) \bigr] \, dx 
          = 1, \; \int_{0}^{\tau} x \bigl[ f(x) + t \rho(x) \bigr] \, dx = \mu; \; \\
            \int_{0}^{\tau} (x - \mu)^{2} \bigl[ f(x) + t \rho(x) \bigr] \, dx = \sigma^{2}, \; \\
              \text{ for all $t$ in a some interval containing } 0.
    \end{multline}
Differentiating \eqref{E:density.constraints} with respect to $t$ we get, 
    \begin{equation} \label{E:rho.constraints}
        \int_{0}^{\tau} \rho(x) \, dx = 0, \; \int_{0}^{\tau} x \rho(x) \, dx = 0; 
          \; \int_{0}^{\tau} (x - \mu)^{2} \rho(x) \, dx = 0.
    \end{equation}
(Let's not worry about the issue of differentiating under the integral sign, shall we?) 

From \eqref{E:rho.constraints} we have
    \begin{align} \label{E:diff.w.r.t..t}
        \frac{\partial H(f + t \rho)}{\partial t}  
          &= - \int_{0}^{\tau} \rho(x) \log \bigl[ f(x) + t \rho(x) \bigr] \, dx 
          - \int_{0}^{\tau} \bigl[ f(x) + t \rho(x) \bigr] \frac{\rho(x)}{f(x) + t \rho(x)} \, dx \notag \\
            &= - \int_{0}^{\tau} \rho(x) \log \bigl[ f(x) + t \rho(x) \bigr] \, dx 
              - \int_{0}^{\tau} \rho(x) \, dx  \\
                &= - \int_{0}^{\tau} \rho(x) \log \bigl[ f(x) + t \rho(x) \bigr] \, dx. \notag 
    \end{align}
For future reference,
\begin{equation*}
        \frac{\partial^{2} H(f + t \rho)}{\partial t^{2}} 
          = - \int_{0}^{\tau} \rho(x) \frac{\rho(x)}{f(x) + t \rho(x)} \, dx.
\end{equation*}
At $t = 0$ this is 
    \begin{equation}  \label{E:2nd.deriv.of.entropy}
        \frac{\partial^{2} H(f + t \rho)}{\partial t^{2}} \vert_{t=0} 
          = - \int_{0}^{\tau} \frac{\rho(x)^{2}}{f(x)} \, dx \leq 0.
    \end{equation}

According to \eqref{E:.partial.H.w.r.t.t.= 0}, if we evaluate the derivative 
$\frac{\partial H(f + t \rho)}{\partial t}$ at $t = 0$ we get 0. From \eqref{E:diff.w.r.t..t} we have,
    \begin{equation*}
        0 = \frac{\partial H(f + t \rho)}{\partial t} \vert_{t=0} = - \int_{0}^{\tau} \rho(x) \log f(x) \, dx.
    \end{equation*}
This has to be true for \emph{any} function $\rho$ satisfying the constraints \eqref{E:rho.constraints}. We can write 
    \begin{equation*}
        \log f(x) = a + b x + c (x - \mu)^{2} + \phi(x), 
    \end{equation*}
where $\phi \in L^{2}[0,\tau]$ is orthogonal to 1, $x$, and $(x - \mu)^{2}$. Thus, \eqref{E:rho.constraints} holds with $\phi$ in place of $\rho$. 
So $\rho = \phi$ is a legitimate choice. Hence, 
    \begin{equation*}
        0 = - \int_{0}^{\tau} \rho(x) \log f(x) \, dx 
          = - \int_{0}^{\tau} \phi(x) \bigl[ a + b x + c (x - \mu)^{2} + \phi(x) \bigr] \, dx 
            = - \int_{0}^{\tau} \phi(x)^{2}(x) \, dx.
    \end{equation*}
Thus, $\phi = 0$ (almost everywhere) and $\log f(x) = a + b x + c (x - \mu)^{2}$. I.e., $f$ is Gaussian, conditional on having support in $[0, \tau]$. Moreover, by \eqref{E:2nd.deriv.of.entropy}, this Gaussian maximizes entropy, $H$. 

The moments $\mu$ and $\sigma^{2}$ in the constraints \eqref{E:moments.of.f} come from data on a random variable whose distribution is approximately absolutely continuous. (``Approximately'' because that subjects' ages are rounded off.) There remains the issue of whether there always exists a Gaussian, truncated to have support in the interval $(0, \tau)$, that satisfies the constraints \eqref{E:moments.of.f}. Apparently, the answer is false. For example, a normal distribution conditioned to have support in a unit interval cannot have an SD larger than that of a uniform over a unit interval, viz., $1/\sqrt{12} \approx 0.287$. We did not find any papers in which the SD was much bigger than that upper bound. In those instances, we fit a conditional Gaussian with SD near the upper bound. 

Given $X \approx Normal(\mu, \sigma^{2})$ the mean, $\mu_{\tau}(\mu, \sigma)$, and variance, $\sigma_{\tau}(\mu, \sigma)^{2}$, of the distribution of $X | \{ 0 < X < \tau \}$ can be computed by numerical integration. The inverse problem of going from observed values or estimates of the conditional moments $(\nu, \omega)$ to the corresponding unconditional can be solved by numerically minimizing the function 
$(\mu, \sigma) \mapsto \Bigl[ \bigl( \mu_{\tau}(\mu, \sigma), \sigma_{\tau}(\mu, \sigma) \bigr) - (\nu, \omega) \Bigr]^{2}$. 

It's useful to have the unconditional moments, for example, for generating simulated ages or computing the conditional density.

Another observation is that if in a study SHB at baseline is measured, not over ``lifetime,'' but over a fixed period, we can approximate it by a random baseline exposure with a very small SD.

\subsubsection{``Default procedure''} \label{SSS:default.procedure}
We can combine the results from the preceding section and section \ref{SSS:gamma.mixture} to formulate a ``default procedure'', one that applies when there are no ``clean'' inclusion/exclusion criteria that need to be taken into account and the baseline proportions are for ``lifetime'', with the understanding that a fixed baseline exposure can be approximated by a random lifetime with very small SD. 

The papers in this category are:
Asarnow \emph{et al} \cite{jrAetAl2017.CBfamTSuiPrevent}, 
Aseltine \emph{et al} \cite{ArhJaSeaGj2007.SOS}, 
Diamond \emph{et al} \cite{gsDetAl.2010.attachment.family}, 
Diamond \emph{et al} \cite{gsDrrKesKEsaLjlHjmRrjG2019.attachment},  
Eggert \emph{et al} \cite{llEeaTbpRkcP2002.SchoolBased}, 
Grupp-Phelan \emph{et al} \cite{jG-PjSsBdmCrtAsL-HkHscMlSjvCjaB2019.Motivational}, 
Hetrick \emph{et al} \cite{seHjGhpYcgDagPjRdjRaMcRlSjGsRmbS2017.Online}, 
Huey \emph{et al} \cite{sjHswHmdRcaH-BpbCsgPjE2004.psychEmergencies}, 
Kennard \emph{et al} \cite{bdKtGaaFdlMcMkWcBaMaDlZeWvOjZsIgPdB2018.ASAP}, 
King \emph{et al} \cite{caKaKlPdcKlWsV2006.YouthNominated}, 
King \emph{et al} \cite{caKnKaKsVpQbG2009.YouthNomSupport}, 
Rengasamy \emph{et al} \cite{mRgS2019.Telephone}, 
and Schilling \emph{et al} \cite{eaSrhAaJ2016.SOS}. 

\subsubsection{Conditioning on there being at least one event in the baseline period} \label{SS:prob.given.at.least.1}
In the study described in Ougrin \emph{et al} \cite{dOiBdSrBeT2013.SuicidePrevention} an inclusion criterion was that a subject must have already performed a self harm behavior (SHB). (See section \ref{SSS:Hazell} concerning the source of the baseline behavior rate we used.)
In the paper we're given the proportion of subjects, all of whom have at least one baseline SHB, who made at least two. This is an estimate of the \emph{conditional} probability of a subject from the population making at least two SHBs in the baseline period \emph{given the event ``made at least 1''}. We need to express this conditional probability in terms of our Cox process model $M_{2}$.

Let $\lambda_{1}$ be the Poisson rate per unit time (one year) that a randomly chosen member of the population would make SHB's in baseline. Use the model from section \ref{SSS:gamma.mixture}, a gamma mixture of Poissons with latent variable 
$R \sim \Gamma(\alpha, \alpha)$. As discussed in subsection \ref{SSS:truncated.normal}, $T_{1}$ is random with a conditional normal distribution. So ultimately, we must take the expectation over $T_{1}$. For now, hold $T_{1}$ fixed.

Let $N_{1}$ be the number of baseline events in $T_{1}$ units of time. Then, by \eqref{E:qi.formla}, 
    \begin{equation*}
       Prob \{ N_{1} \geq 1 | T_{1} \} = 1 -  \exp(- R \lambda_{1} T_{1}) 
         = 1 - \left( \frac{\alpha}{\alpha + \lambda_{1} T_{1}} \right)^{\alpha} 
    \end{equation*}
 
Similarly, the conditional probability given $T_{1}$ of making at least two SHBs is 
    \begin{equation*}
     1 - E \exp(- R \lambda_{1} T_{1}) 
       - E \bigl[ R \lambda T_{1} \exp(- R \lambda_{1} T_{1}) \bigr].
    \end{equation*}
By \eqref{E:qi.formla} and \eqref{E:unconditional.prob.of.1}, this equals
    \begin{equation}
      1 - \left( \frac{\alpha}{\alpha + \lambda_{1} T_{1}} \right)^{\alpha} 
        -  \lambda_{1} T_{1} \left( \frac{\alpha}{\alpha + \lambda_{1} T_{1} } \right)^{\alpha+1} .
    \end{equation}
Hence, the conditional probability given $T_{1}$ of making at least two SHBs given having made at least one is 
    \begin{equation} \label{E.conditional.prob.of.making.more.than.1.attempt}
        1 - \frac{ \left( \frac{\alpha}{\alpha + \lambda_{1} T_{1} } \right)^{\alpha+1} }
          { 1 - \left( \frac{\alpha}{\alpha + \lambda_{1} T_{1}} \right)^{\alpha}  } . 
    \end{equation}
We set this equal to the observed conditional probability and solve for $\lambda_{1}$.

Turn now to McCauley \emph{et al} \cite{eMcMEtal.2018.DBTforSuicidalBehavior}. 
Given at least one event in baseline, what's the probability of an event in follow-up? Let $N_{2}$ be the number of follow-up SHBs in $T_{2}$ units of time. Then, as in \eqref{E:qi.formla},
    \begin{align*}
      Prob \{ N_{2} = 0 \text{ and } N_{1} > 0 \} 
        &= E \Bigl[ \exp \{ - R \lambda_{2} T_{2} \} (1 - \exp \{ - R \lambda_{1} T_{1} \}) \Bigr] \\
        &= E \Bigl( \exp \{ - R \lambda_{2} T_{2} \} 
          - \exp \{ - R (\lambda_{1} T_{1} + \lambda_{2} T_{2}) \} \Bigr) \\
        &= \left( \frac{\alpha}{\alpha + \lambda_{2} T_{2}} \right)^{\alpha} 
           - \left( \frac{\alpha}{\alpha + \lambda_{1} T_{1} + \lambda_{2} T_{2})} \right)^{\alpha} .
    \end{align*}
Therefore,
    \begin{equation}  \label{E:poz.in.flup.given.poz.at.base}
      Prob \{ N_{2} > 0 | N_{1} > 0 \} 
        = 1 - \frac{ \left( \frac{\alpha}{\alpha + \lambda_{2} T_{2}} \right)^{\alpha} 
               - \left( \frac{\alpha}{\alpha + \lambda_{1} T_{1} + \lambda_{2} T_{2})} \right)^{\alpha} }
                 { 1 - \left( \frac{\alpha}{\alpha + \lambda_{1} T_{1}} \right)^{\alpha}  } .
    \end{equation}

\subsubsection{Conditioning on there being at least two events in the baseline period with one ``at'' enrollment}  \label{SS:expected.number.given.at least.2}
In the study described in Wood \cite{aWgTjRaMrH.TreatmentFollowingSelfHarm}, an inclusion criterion was that a subject has just committed an act of self-harm and had prior incident within the last year. 
If the exposure time is random, (1) what is the expected number of events that take place lifetime (so far)? And (2), given this baseline behavior, what's the expected number of follow-up events? (Cottrell \cite{djCEtal.2018.FamilyTherapy} is similar. See below.)

Let $\alpha > 0$, $\beta \geq 0$ and $\lambda_{1} > 0$ and define
    \begin{equation} \label{E:f.beta.defn}
      f_{\beta}(s) := f_{\lambda_{1}, \beta}(s) 
        := \left( \frac{\alpha}{\alpha + \lambda_{1} s } \right)^{\alpha+\beta}, 
          \quad s > - \alpha/\lambda_{1} < 0  .
\end{equation}
Thus, $f_{\beta}(0) = 1$ and
    \begin{equation} \label{E:f.beta.deriv}
      f_{\beta}'(s) := - \lambda_{1} \frac{\alpha+\beta}{\alpha}
        \left( \frac{\alpha}{\alpha + \lambda_{1} s } \right)^{\alpha+\beta+1}
          =   - \lambda_{1} \frac{\alpha+\beta}{\alpha} f_{\beta+1}(s),
            \quad s > - \alpha/\lambda_{1} < 0  .
    \end{equation}
Applying \eqref{E:f.beta.deriv} recursively we get 
    \begin{multline} \label{E:f.beta.2nd.deriv}
      f_{\beta}''(s) 
        = \lambda_{1}^{2} \frac{\alpha+\beta}{\alpha} \frac{\alpha+\beta + 1}{\alpha} f_{\beta+2}(s) \\
         = \lambda_{1}^{2} \frac{\alpha^{\alpha+\beta}(\alpha+\beta)(\alpha+\beta + 1)}
           {(\alpha + \lambda_{1} s)^{\alpha+\beta+2} }  ,
            \quad s > - \alpha/\lambda_{1} < 0  .
    \end{multline}

Suppose $s > 0$ and, given the $\Gamma(\alpha, \alpha)$ variable $R$, $N$ is Poisson 
with mean $\lambda s$ As in \eqref{E:qi.formla} we have
    \begin{equation}    \label{E:unconditional.prob.of.0}
        Prob \{ N = 0 \} = E (e^{-\lambda s R})
          = \left( \frac{\alpha}{\alpha + \lambda s} \right)^{\alpha}
           = f_{\lambda, 0}(s) .
    \end{equation}
Then, by \eqref{E:f.beta.deriv},  
    \begin{equation}  \label{E:unconditional.prob.of.1}
        Prob \{ N = 1 \} = E (\lambda s R e^{-\lambda s R}) = - s f_{\lambda, 0}' (s) 
          = \lambda s \left( \frac{\alpha}{\alpha + \lambda s} \right)^{\alpha+1}
            = \lambda s f_{\lambda, 1}(s) .
    \end{equation}
And by \eqref{E:f.beta.2nd.deriv},  
    \begin{multline}   \label{E:unconditional.prob.of.2}
        Prob \{ N = 2 \} = E \left( \frac{ (\lambda s R)^{2} e^{-\lambda s R} }{2} \right)
         = \frac{s^{2}}{2} f_{\lambda, 0}'' (s)   \\
          = \frac{\lambda^{2} s^{2} \alpha^{\alpha+1}}{2 (\alpha + \lambda s)^{\alpha+2}} \, (\alpha+1) 
            = \frac{\lambda^{2} s^{2} (\alpha + 1)}{2 \alpha} f_{\lambda, 2}(s) .
    \end{multline}
(See \eqref{E:general.R^q.exp(-lambda.R).formla} and \eqref{E:probs.in.terms.of.f}.)

Let $R$ be our usual gamma random variable. Let $\lambda_{1} > 0$ and $s > 0$ be non-random. Given $R$ let $N$ be Poisson with mean $\lambda_{1} s R$. By \eqref{E:qi.formla}, \eqref{E:general.R^q.exp(-lambda.R).formla}, and \eqref{E:f.beta.defn} we have
    \begin{align} \label{E:probs.in.terms.of.f}
      Prob\{ N > 0 \} &= 1 - f_{0}(s) , \\
      Prob\{ N = 2 \} &= 
        \frac{(\lambda_{1} s)^{2}}{2} E \bigl[ R^{2} \exp(-\lambda_{1} s R) \bigr] 
        = \frac{(\lambda_{1} s)^{2}}{2} \frac{\alpha+1}{\alpha+\lambda_{1}s} f_{1}(s) . 
          \notag
    \end{align}
(See \eqref{E:unconditional.prob.of.1} and \eqref{E:unconditional.prob.of.2}.)

Suppose, given the gamma distributed latent variable $R$, events at baseline follow a Poisson process with rate $R \lambda_{1}$. Let $t > \Delta > 0$, e.g., $\Delta =$ one year. Let $M$ be the number of events in $[0, t]$ and suppose there was an event at time $t$. Include this point in $M$. Let $\tilde{M}$ be the number of events in 
the interval $[0, t - \Delta)$. Given $R$, $\tilde{M}$ is Poisson with conditional mean 
$R \lambda_{1} (t - \Delta)$.  Let $M'$ be the number of events in the interval 
$[t - \Delta, t)$, so excluding $\{t\}$.
Then, given $R$, $M'$ is Poisson with conditional mean $R \lambda_{1} \Delta$. 
Thus, given $R$, $\tilde{M}+M'$ is Poisson with mean $R \lambda_{1} t$.

What's the conditional expectation of $M = \tilde{M}+M'+1$ given $R$, $\{ M' > 0 \}$, and
an event ``at'' $t$? Let $\delta \in (0, \Delta)$ be small. Then with high probability, the number, $M'_{\delta}$, of events in $[t - \Delta, t-\delta]$ will equal $M'$ and with high probability the number, $M_{\delta}$, of events in $(t-\delta, t]$ 
will be $\leq 1$, in fact 0. A more precise estimate is 
$E (M_{\delta} | R ) \approx Prob \{ M_{\delta} > 0 | R \} \approx \lambda R \delta$.
We have $M = \tilde{M}+M'_{\delta}+M_{\delta}$. Therefore, what we want is approximately 
$E (M | R, M'_{\delta} > 0, M_{\delta} > 0)/(\lambda R \delta)$. We work this out now. 
Let $u := t-\Delta$. We have
    \begin{multline} \label{E:EM.M'.delta>0.M.delta>0.decomp}
      E \bigl( M 1_{\{M'_{\delta} > 0\}} 1_{\{M_{\delta} > 0\}} | R \bigr)  \\
        = E \bigl( \tilde{M} 1_{\{M'_{\delta} > 0\}} 1_{\{M_{\delta} > 0\}} | R \bigr)
          + E \bigl( (M'_{\delta} + M_{\delta}) 1_{\{M'_{\delta} > 0\}} 
            1_{\{M_{\delta} > 0\}} | R \bigr) .
    \end{multline}

First, analyze $E \bigl( (M'_{\delta} + M_{\delta}) 1_{\{M'_{\delta} > 0\}} 1_{\{M_{\delta} > 0\}} | R \bigr)$. The following is obvious, but one might appeal also to the theorem in 
Feller \cite[p.\ 89]{wF57.feller1} 
    \begin{align*}
      E \bigl( (M'_{\delta} &+ M_{\delta}) 1_{\{M'_{\delta} > 0\}} 
        1_{\{M_{\delta} > 0\}} | R \bigr) \\
      &= E ( M'_{\delta} + M_{\delta} | R )
      - E \bigl( (M'_{\delta} + M_{\delta}) 1_{\{M'_{\delta} = 0\}} | R \bigr) \\
      & \qquad - E \bigl( (M'_{\delta} + M_{\delta}) 1_{\{M_{\delta} = 0\}} | R \bigr) \\
      & \qquad + E \bigl( (M'_{\delta} + M_{\delta}) 1_{\{M'_{\delta} = 0\}} 1_{\{M_{\delta} = 0\}} 
        | R \bigr) \\
        &= E ( M'_{\delta} + M_{\delta} | R ) - E ( M_{\delta} 1_{\{M'_{\delta} = 0\}} | R )
          - E ( M'_{\delta} 1_{\{M_{\delta} = 0\}} | R ) \\
            &= \lambda_{1} R \Delta - \lambda_{1} R \delta 
              \exp(- \lambda_{1} R (\Delta - \delta) ) 
              - \lambda_{1} R (\Delta-\delta) \exp(- \lambda_{1} R \delta ) .
    \end{align*}

Therefore, by \eqref{E:ER.=.1}, \eqref{E:general.R^q.exp(-lambda.R).formla}, \eqref{E:f.beta.defn}, and the fact that $f_{\beta}(0) = 1$, we have,
    \begin{multline}  \label{E:E(M'.delta+M.delta|M'.delta.>.0.and.M.delta.>.0)}
      E \bigl( (M'_{\delta} + M_{\delta}) 1_{\{M'_{\delta} > 0\}} 1_{\{M_{\delta} > 0\}} \bigr)  \\
        = \lambda_{1} \Delta - \lambda_{1} \delta \left( \frac{\alpha}{\alpha 
          + \lambda_{1} (\Delta - \delta)} \right)^{\alpha+1} 
              - \lambda_{1} (\Delta-\delta) 
                \left( \frac{\alpha}{\alpha + \lambda_{1} \delta} \right)^{\alpha+1} \\
                  = \lambda_{1} \Delta - \lambda_{1} \delta f_{1}(\Delta - \delta) 
                    - \lambda_{1} (\Delta-\delta) f_{1}(\delta) \\
                      = \lambda_{1} \left[ - \Delta \bigl( f_{1}(\delta) - f_{1}(0) \bigr) -
                        \delta \bigl( f_{1}(\Delta-\delta) - f_{1}(\delta) \bigr) \right] .
    \end{multline}
Thus, by the Mean Value Theorem and \eqref{E:f.beta.deriv},  
there exists $\delta' \in [0, \delta]$ s.t.
    \begin{align} \label{E:EM'.delta.+.M.delta.post.MVT}
      E \bigl( (M'_{\delta} &+ M_{\delta}) 1_{\{M'_{\delta} > 0\}} 1_{\{M_{\delta} > 0\}} \bigr)
          \notag \\
        &= \lambda_{1} \delta \left[ - \Delta f_{1}'(\delta') - f_{1}(\Delta-\delta) 
          + f_{1}(\delta) \bigr) \right] \\
        &= \lambda_{1} \delta \left[ \Delta \lambda_{1} \frac{\alpha+1}{\alpha} f_{2}(\delta') -
               f_{1}(\Delta-\delta) + f_{1}(\delta) \bigr) \right] \notag .
    \end{align}

In a similar way we analyze $E \bigl( \tilde{M} 1_{\{M'_{\delta}  > 0\}} 1_{\{M_{\delta} > 0\}} | R \bigr)$:
    \begin{multline*}  
      E \bigl( \tilde{M} 1_{\{M'_{\delta} > 0\}} 1_{\{M_{\delta} > 0\}} | R \bigr) \\
        = \lambda_{1} R u \bigl[ 1 - \exp(- \lambda_{1} (\Delta-\delta) R) \bigr] 
          \bigl[ 1 - \exp(- \lambda_{1} \delta R) \bigr] \\
            = \lambda_{1} u \bigl[ R - R \exp(- \lambda_{1} (\Delta-\delta) R) 
              -  R \exp(- \lambda_{1} \delta R)  \\
                + R \exp(- \lambda_{1} \Delta R)  \bigr] .
    \end{multline*}
Therefore, \eqref{E:ER.=.1}, \eqref{E:probs.in.terms.of.f}, and the fact that
$f_{\beta}(0) = 1$, we have, 
    \begin{multline*}  
      E \bigl( \tilde{M} 1_{\{M'_{\delta} > 0\}} 1_{\{M_{\delta} > 0\}} \bigr) 
        = \lambda_{1} u \bigl[ 1 - f_{1}(\Delta-\delta) - f_{1}(\delta) + f_{1}(\Delta) \bigr] \\
          =  \lambda_{1} u \Bigl( \bigl[f_{1}(\Delta) - f_{1}(\Delta-\delta) \bigr] 
            - \bigl[ f_{1}(\delta) - f_{1}(0) \bigr] \Bigr) .
    \end{multline*}
Hence, by the Mean Value Theorem and \eqref{E:f.beta.deriv}, 
there exists $\delta'', \delta''' \in [0, \delta]$ s.t.
    \begin{equation*}  
      E \bigl( \tilde{M} 1_{\{M'_{\delta} > 0\}} 1_{\{M_{\delta} > 0\}} \bigr)
          =  \lambda_{1}^{2} u \delta \, \frac{\alpha+1}{\alpha} \bigl[ - f_{2}(\Delta - \delta'') + f_{2}(\delta''') \bigr] .
    \end{equation*}
 Combining this with \eqref{E:EM.M'.delta>0.M.delta>0.decomp} and \eqref{E:EM'.delta.+.M.delta.post.MVT}, we get
     \begin{multline*}  
       E \bigl[ M 1_{\{M'_{\delta} > 0\}} 1_{\{M_{\delta} > 0\}} \bigr] \\
          =  \lambda_{1} \delta \Bigl( \lambda_{1} u \frac{\alpha+1}{\alpha} \bigl[ - f_{2}(\Delta - \delta'') 
            + f_{2}(\delta''') \bigr] \\
            + \Delta \lambda_{1} \frac{\alpha+1}{\alpha} f_{2}(\delta') -
               f_{1}(\Delta-\delta) + f_{1}(\delta) \bigr) \Bigr) .
    \end{multline*}
 
 Therefore, by \eqref{E:Prob.M'.delta>0.and.M.delta>0}, we have 
     \begin{multline*}  
       E \bigl[ M | M'_{\delta} > 0, \, M_{\delta} > 0 \bigr] \\
          =  \Bigl( \lambda_{1} u \frac{\alpha+1}{\alpha} \bigl[ - f_{2}(\Delta - \delta'') 
            + f_{2}(\delta''') \bigr] \\
            + \Delta \lambda_{1} \frac{\alpha+1}{\alpha} f_{2}(\delta') -
               f_{1}(\Delta-\delta) + f_{1}(\delta) \bigr) \Bigr) \div
                 \Bigl( f_{1}(\epsilon')  - f_{1}(\Delta - \epsilon'')  \bigr ) \Bigr) .
    \end{multline*}
 
 Now, $u+\Delta = t$ so, letting $\delta \downarrow 0$, we get
      \begin{multline}  \label{E:E(M'.0+0|M'.0.>.0.and.M.0.>.0}
       E \bigl[ M | M'_{0^{+}} > 0, \, M_{0^{+}} > 0 \bigr] \\
          =  \Bigl( \lambda_{1} u \frac{\alpha+1}{\alpha} \bigl[ - f_{2}(\Delta) 
            + f_{2}(0) \bigr] \\
            + \Delta \lambda_{1} \frac{\alpha+1}{\alpha} f_{2}(0) -
               f_{1}(\Delta) + f_{1}(0) \bigr) \Bigr) \div
                 \Bigl( f_{1}(0)  - f_{1}(\Delta) \Bigr) \\
                 = \Bigl( \lambda_{1} t \frac{\alpha+1}{\alpha} 
                   - \lambda_{1} u \frac{\alpha+1}{\alpha} f_{2}(\Delta) - 
               f_{1}(\Delta) +1 \bigr) \Bigr) \div
                 \Bigl( 1  - f_{1}(\Delta)  \bigr ) \Bigr) \\
                 = \lambda_{1} \frac{\alpha+1}{\alpha} 
                   \times \frac{ t - u  f_{2}(\Delta)}{ 1  - f_{1}(\Delta) } + 1 .
    \end{multline}

Now let $N$ be the number of SHBs that occur in a disjoint interval of length $T$ (think follow-up) with a Poisson rate of $\lambda_{2}$. We want the conditional expectation of $N$ given 
$M' > 0$ \emph{and} and there is an event at time $\Delta$. Do an informal calculation. Let $\delta > 0$ be small. Then with high probability, the number, $M'_{\delta}$, of events in $[0, \Delta-\delta]$ will equal $M'$ and with high probability the number, $M_{\delta}$, of events in $(\Delta-\delta, \Delta]$ will be $\leq 1$, proportional to $\delta$ on average. Therefore, what we want is approximately $E (N | M'_{\delta} > 0, M_{\delta} > 0)$. We have
    \begin{multline*}
      E \bigl( N 1_{\{M'_{\delta} > 0\}} 1_{\{M_{\delta} > 0\}} | R \bigr) = 
        R \lambda_{2} T \bigl( 1 - \exp(-R \lambda_{1} (\Delta-\delta)) \bigr) 
          \bigl( 1 - \exp(-R \lambda_{1}  \delta) \bigr) \\
            =  \lambda_{2} T \Bigl( R - R \exp(-R \lambda_{1}  \delta) 
              - R \exp \bigl( -R \lambda_{1} (\Delta-\delta) \bigr)
                + R \exp(-R \lambda_{1} \Delta) \Bigr) .
    \end{multline*}
Therefore, by \eqref{E:general.R^q.exp(-lambda.R).formla},  
    \begin{multline*}
      (\lambda_{2} T)^{-1} E \bigl( N 1_{\{M'_{\delta} > 0\}} 1_{\{M_{\delta} > 0\}} \bigr) \\  
        = E \Bigl( R - R \exp(-R \lambda_{1}  \delta) 
          - R \exp \bigl( -R \lambda_{1} (\Delta-\delta) \bigr) 
            + R \exp(-R \lambda_{1} \Delta) \Bigr) \\
              =  1 - \left( \frac{\alpha}{\alpha + \lambda_{1} \delta } \right)^{\alpha+1} 
                - \left( \frac{\alpha}{\alpha + \lambda_{1} (\Delta-\delta) } \right)^{\alpha+1}
                  + \left( \frac{\alpha}{\alpha + \lambda_{1} \Delta } \right)^{\alpha+1} .
    \end{multline*}

Now, $f_{\beta}(0) = 1$. Hence, 
    \begin{equation*}
      (\lambda_{2} T)^{-1} E \bigl( N 1_{\{ M'_{\delta} > 0\}} 1_{\{M_{\delta} > 0 \} } \bigr)
       = - \bigl[ f_{1}(\delta) - f_{1}(0) \bigr] + \bigl[ f_{1}(\Delta) - f_{1}(\Delta-\delta) \bigr]  . 
    \end{equation*}
Now, by \eqref{E:f.beta.deriv}, 
    \begin{equation} \label{E:f1.deriv}
      f_{1}'(s) := - \lambda_{1} \frac{\alpha+1}{\alpha } f_{2}(s) . 
    \end{equation}
Therefore, by the Mean Value Theorem and \eqref{E:f1.deriv}, 
for some $\delta', \delta'' \in [0, \delta]$,
    \begin{equation} \label{E:EN.M'.M}
      (\lambda_{2} T)^{-1} E \bigl( N 1_{\{M'_{\delta} > 0\}} 1_{\{M_{\delta} > 0\}} \bigr)
       = \delta \lambda_{1} \frac{\alpha+1}{\alpha} f_{1}(\delta') 
         - \delta \lambda_{1} \frac{\alpha+1}{\alpha} f_{1}(\Delta - \delta'') . 
    \end{equation}

Similarly, we have, 
    \begin{multline*}  
      Prob \{ M'_{\delta} > 0 \text{ and } M_{\delta} > 0 | R \} 
       = \bigl( 1 - \exp(-R \lambda_{1} (\Delta-\delta)) \bigr) \bigl( 1 - \exp(-R \lambda_{1}  \delta) \bigr) \\
         = 1 - \exp(-R \lambda_{1}  \delta) - \exp \bigl( -R \lambda_{1} (\Delta-\delta) \bigr) + \exp(-R \lambda_{1} \Delta) .
    \end{multline*}
Therefore, by \eqref{E:qi.formla}, 
    \begin{multline} \label{E:Prob.M'.M}
      Prob \{ M'_{\delta} > 0 \text{ and } M_{\delta} > 0 \} \\
         = E \left[ 1 - \exp(-R \lambda_{1}  \delta) - \exp \bigl( -R \lambda_{1} (\Delta-\delta) \bigr) 
            + \exp(-R \lambda_{1} \Delta) \right] \\
              = - \bigl[ f_{0}(\delta) - f_{0}(0) \bigr] + \bigl[ f_{0}(\Delta) - f_{0}(\Delta-\delta) \bigr]
    \end{multline}

Now, by \eqref{E:f.beta.deriv},
    \begin{equation}  \label{E:f.0.deriv}
      f_{0}'(s) = - \lambda_{1} f_{1}(s)  .
    \end{equation} 
Therefore, by the Mean Value Theorem again 
for some $\epsilon', \epsilon'' \in [0, \delta]$,
    \begin{equation}  \label{E:Prob.M'.delta>0.and.M.delta>0}
      Prob \{ M'_{\delta} > 0 \text{ and } M'_{\delta} > 0 \}  
        \approx \delta \lambda_{1} f_{1}(\epsilon')  - \delta \lambda_{1} f_{1}(\Delta - \epsilon'')  .
    \end{equation}

Hence, by \eqref {E:EM'.delta.+.M.delta.post.MVT} for some $\epsilon', \epsilon'' \in [0, \delta]$,
    \begin{equation*} \label{E: E((M'.delta.+M.delta.|.M'.delta.>.0.and.M.delta.>.0)}
      E \bigl( M | M'_{\delta} > 0 \text{ and } M_{\delta} > 0 \bigr) 
        \approx \frac{ \Delta \lambda_{1} \frac{\alpha+1}{\alpha} f_{2}(\delta') + f_{1}(\delta) -
           f_{1}(\Delta-\delta)}{  f_{1}(\epsilon') - f_{1}(\Delta-\epsilon'') } .
    \end{equation*}
Therefore, letting $\delta \downarrow 1$, we get finally,
    \begin{equation} \label{E:lE((M'.delta.+M.delta.|.M'.delta.>.0.and.M.delta.>.0)}
      E \bigl( (M'_{0^{+}} + M_{0^{+}}) | M'_{0^{+}} > 0 \text{ and } M_{0^{+}} > 0 \bigr) 
          = \frac{ (\alpha+1) \lambda_{1} \Delta }{  \alpha(1 - f_{1}(\Delta) ) } + 1 . 
    \end{equation}
 
Similarly, dividing \eqref{E:EN.M'.M} by \eqref{E:Prob.M'.delta>0.and.M.delta>0} 
and letting $\delta \downarrow 0$, we get
    \begin{equation*}
     E (N | M'_{0^{+}} > 0, M_{0^{+}} > 0) 
                 = \lambda_{2} T \, \frac{\alpha+1}{\alpha} \, \frac{ 1 -  f_{1}(\Delta) }{ 1 -  f_{1}(\Delta)  } 
                   = \lambda_{2} T \,  \frac{\alpha+1}{\alpha} .
    \end{equation*}

Now compute similar conditional probabilities, ones from Cottrell \cite{djCEtal.2018.FamilyTherapy}. First, we want to approximate 
$Prob \{ M'_{\delta} + M_{\delta} > 2 | M'_{\delta} > 0 \text{ and } M_{\delta} > 0 \}$. We have
    \begin{multline*}
     Prob \{ M'_{\delta} + M_{\delta} > 2, M'_{\delta} > 0, \text{ and } M_{\delta} > 0 \} \\
       = Prob \{ M'_{\delta} > 0 \text{ and } M_{\delta} > 1 \} + Prob \{ M'_{\delta} > 1 \text{ and } M_{\delta} = 1 \}.
    \end{multline*}

Now $Prob \{ M'_{\delta} > 0 \text{ and } M_{\delta} > 1 \} \leq Prob \{ M_{\delta} > 1 \}$. We \emph{claim}
     \begin{equation}  \label{E:Prob.Mdelta>1.=o(delta)}
      Prob \{ M_{\delta} > 1 \} = o(\delta) \text{  as } \delta \downarrow 0,
    \end{equation}
where we employ Landau ``big and little o'' notation, de Bruijn \cite[Section 1.3]{ngdeB81.AsympMthdsAnlys}. By \eqref{E:probs.in.terms.of.f},
    \begin{align*}
      Prob \{ M_{\delta} > 1 \} / \delta 
       &= \delta^{-1} E Prob \{ M_{\delta} > 1 | R \} \\
       &= \delta^{-1} \Bigl( 1 - Prob \{ M_{\delta} = 0 \}  -  Prob \{ M_{\delta} = 1 \} \Bigr) \\
       &= \delta^{-1} \Bigl( 1 - E \, \exp (- \lambda_{1} \delta R ) 
         -  \lambda_{1} \delta E \, R \exp (- \lambda_{1} \delta R  \Bigr) \\
       &= \delta^{-1} \bigl( 1 - f_{0}(\delta) -  \lambda_{1} \delta f_{1}(\delta) \bigr) \\
       &= \delta^{-1} \bigl( 1 - f_{0}(\delta) \bigr) -  \lambda_{1} f_{1}(\delta) .
    \end{align*}
By the Mean Value Theorem and \eqref{E:f.beta.deriv} there exists 
$\tilde{\delta} \in [0, \delta]$ s.t.\
    \begin{multline*}
      Prob \{ M_{\delta} > 1 \} / \delta 
        =  \delta^{-1} \bigl( 1 - f_{0}(\delta) \bigr) -  \lambda_{1} f_{1}(\delta) \\
          = - f_{0}'(\tilde{\delta}) -  \lambda_{1} f_{1}(\delta)
            = \lambda_{1} f_{1}(\tilde{\delta}) - \lambda_{1} f_{1}(\delta) \to 0 
              \text{ as } \delta \downarrow 0 .
    \end{multline*}
This proves the claim that $Prob \{ M_{\delta} > 1 \} = o(\delta)$. Thus,
    \begin{multline}  \label{E:M'delta+M'delta>2.M'delta>0.Mdelta> 0} 
      Prob \{ M'_{\delta} + M_{\delta} > 2, M'_{\delta} > 0, \text{ and } M_{\delta} > 0 \} \\
        = o(\delta) + Prob \{ M'_{\delta} > 1, \text{ and } M_{\delta} = 1 \} .
    \end{multline}
Write $t$ instead of $\Delta$ now. By \eqref{E:probs.in.terms.of.f}, we have
    \begin{align*}
      Prob \{ M'_{\delta} > &1 \text{ and } M_{\delta} = 1 \} \\
        &= E Prob \{ M'_{\delta} > 1 \text{ and } M_{\delta} = 1 | R \} \\
        &= E \Bigl( Prob \{ M'_{\delta} > 1 | R \} Prob \{ M_{\delta} = 1 | R \} \Bigr) \\
        &= E \Bigl( \bigl[ 1 - \exp (- \lambda_{1} (t-\delta) R) 
           - \lambda (t-\delta) R \exp (- \lambda_{1} (t-\delta) R) \bigr]  \\
        & \qquad \qquad \lambda_{1} \delta R \exp (- \lambda_{1} \delta R) \Bigr) \\
        &= \lambda_{1} \delta \, E \bigl[ R \exp (- \lambda_{1} \delta R) - R \exp (- \lambda_{1} t R) 
           - \lambda (t-\delta) R^{2} \exp (- \lambda_{1} t R) \bigr]  \\
        &= \lambda_{1} \delta \,\left[ f_{1}(\delta)- f_{1}(t) 
           - \lambda (t-\delta) f_{1}(t)  \frac{\alpha+1}{\alpha + \lambda_{1} t } \right]  \\
        &= \lambda_{1} \delta \, \left[ f_{1}(\delta) 
          - \frac{\alpha + 2 \lambda_{1} t + \lambda_{1} \alpha t + O(\delta)}
            {\alpha + \lambda_{1} t } f_{1}(t) \right] .
    \end{align*}

Therefore, by \eqref{E:Prob.M'.delta>0.and.M.delta>0}, \eqref{E:Prob.Mdelta>1.=o(delta)}, and \eqref{E:M'delta+M'delta>2.M'delta>0.Mdelta> 0}, we have 
    \begin{multline*}
      Prob \{ M'_{\delta} + M_{\delta} > 2  | M'_{\delta} > 0 \text{ and } M_{\delta} > 0 \} \\
        = \frac{ o(1) +  f_{1}(\delta) -  \frac{\alpha + 2 \lambda_{1} t + \lambda_{1} \alpha t 
          + O(\delta)}{\alpha + \lambda_{1} t } f_{1}(t)}{ f_{1}(\epsilon')  - f_{1}(t - \epsilon'') } ,
    \end{multline*}
where $\epsilon', \epsilon'' \in [0, \delta]$. Finally, letting $\delta \downarrow 0$, we get the approximation,
\begin{equation*}
 Prob \{ M'_{\delta} + M_{\delta} > 2  | M'_{\delta} > 0 \text{ and } M_{\delta} > 0 \}
   \approx \frac{ 1 - \frac{\alpha + 2 \lambda_{1} t + \lambda_{1} \alpha t }
     {\alpha + \lambda_{1} t } f_{1}(t) }{ 1 - f_{1}(t) } .
\end{equation*}

Now, let $N$ be the number of events in an interval of length $T$ disjoint from the interval $[0, t]$. Given $R$, $N$ is Poisson with rate $\lambda_{2}$ events per unit time. We are interested in approximating the conditional probability
$Prob \{ N > 0 | M'_{\delta} > 0, \text{ and } M_{\delta} > 0 \}$. By \eqref{E:Prob.Mdelta>1.=o(delta)}, we have
     \begin{align}  \label{E:Prob.N.M'delta.Mdelta.o}
      Prob \{ N > 0 , M'_{\delta}& > 0 , \text{ and } M_{\delta} > 0 \} \notag \\
        &= Prob \{ N > 0, M'_{\delta} > 0, \text{ and } M_{\delta} > 1 \}  \\
        & \qquad \qquad + Prob \{ N > 0, M'_{\delta} > 0, \text{ and } M_{\delta} = 1 \} \notag \\
        &= o(\delta) + Prob \{ N > 0, M'_{\delta} > 0, \text{ and } M_{\delta} = 1 \} . \notag 
     \end{align}
We have 
    \begin{align*}
        E Prob \{ N > 0,& \; M'_{\delta} > 0, \text{ and } M_{\delta} = 1 | R \} \\
        &= E \Bigl( Prob \{ N > 0 | R \}  Prob \{ M'_{\delta} > 0 | R \} 
          Prob \{ M_{\delta} = 1 | R \} \Bigr) \\
        &= E \Bigl( \bigl[ 1 - \exp (- \lambda_{2} R T) \bigr] 
          \bigl[ 1 - \exp (- \lambda_{1} (t-\delta) R) \bigr]  \\
        & \qquad \qquad \lambda_{1} \delta R \exp (- \lambda_{1} \delta R) \Bigr) \\
        &= \lambda_{1} \delta \, E \bigl[ R \exp (- \lambda_{1} \delta R) 
          - R \exp (- \lambda_{1} t R) \\
        & \qquad \qquad - R \exp \bigl( - (\lambda_{1} \delta + \lambda_{2} T) R \bigr) 
            + R \exp \bigl( - (\lambda_{1} t + \lambda_{2} T) R \bigr) \bigr]  \\
        &= \lambda_{1} \delta \,\left[ f_{1}(\delta)- f_{1}(t) 
          - f_{\lambda_{1} \delta + \lambda_{2} T, 1}(1)
            - f_{\lambda_{1} t + \lambda_{2} T, 1}(1) \right]  .
    \end{align*}
Therefore, by \eqref{E:Prob.N.M'delta.Mdelta.o} and \eqref{E:Prob.M'.delta>0.and.M.delta>0},
    \begin{multline*}
        Prob \{ N > 0 | M'_{\delta} > 0, \text{ and } M_{\delta} > 0 \} \\
        = \frac{o(1) +  f_{1}(\delta)- f_{1}(t) - f_{\lambda_{1} \delta + \lambda_{2} T, 1}(1)
            - f_{\lambda_{1} t + \lambda_{2} T, 1}(1) }{f_{1}(\epsilon')  - f_{1}(t - \epsilon'')} ,
    \end{multline*}
where $\epsilon', \epsilon'' \to 0$ as $\delta \downarrow 0$. Then letting $\delta \downarrow 0$ we get the approximation
    \begin{equation*}
        Prob \{ N > 0 | M'_{\delta} > 0, \text{ and } M_{\delta} > 0 \} 
          \approx \frac{ 1 - f_{1}(t) - f_{\lambda_{2} T, 1}(1) 
            + f_{\lambda_{1} t + \lambda_{2} T, 1}(1) } {1  - f_{1}(t)} .
    \end{equation*}

The right had side expands out at follows:
\begin{equation*}
    \frac{ 1 
      - \left( \tfrac{\alpha}{\alpha + \lambda_{1} t } \right)^{\alpha+1} 
        - \left( \tfrac{\alpha}{\alpha + \lambda_{2} T } \right)^{\alpha+1} 
          + \left( \tfrac{\alpha}{\alpha + \lambda_{1} t + \lambda_{2} T } \right)^{\alpha+1} }
            { 1  - \left( \tfrac{\alpha}{\alpha + \lambda_{1} t + \lambda_{2} T } \right)^{\alpha+1} }
\end{equation*}

\subsubsection{Conditioning on there being an event ``at'' enrollment}  \label{SS:condition.on.1.at.enrollment}
Now we analyze the study described in Donaldson \cite{dDaScE-S05.TreatmentFollowingAttempt}. (See section \ref{SS:prob.given.at.least.1}.) What's the probability that there will be at least 2 events in a baseline interval of length( $t$, given that there's an event ``at'' t? By ``at t'' we will mean in a short interval $[t-\delta, t]$. Let $M^{\delta}$ be the number of events in the interval $[0, t-\delta)$ and let $M_{\delta}$ be the number of events in $[t-\delta, t]$. We have
    \begin{multline} \label{E:M^delta+M_delta.decomp}
          Prob \{ M^{\delta} + M_{\delta} > 1, \, M_{\delta} > 0 \}  \\
            = Prob \{ M^{\delta} > 0, \,  M_{\delta} > 0 \} + Prob \{ M^{\delta} = 0, \,  M_{\delta} > 1 \} .
    \end{multline}
Now, by \eqref{E:Prob.M'.delta>0.and.M.delta>0} with $\Delta = t$ we get 
    \begin{equation} \label{E:M^delta>0.M_delta>0}
          Prob \{ M^{\delta} > 0, \,  M_{\delta} > 0 \} 
            = \delta \lambda_{1} \bigl[ f_{1}(\epsilon')  -  f_{1}(t - \epsilon'') \bigr], 
    \end{equation}
where $f_{\beta}$ is defined in \eqref{E:f.beta.defn} and $\epsilon, \epsilon' \in [0, \delta]$. 

Now we compute 
    \begin{multline*}
         Prob \{ M^{\delta} = 0, \,  M_{\delta} > 1 \}  \\
           = E \Bigl( \exp( - \lambda_{1} (t-\delta) R) 
             \bigl[ 1 - \exp(-\lambda_{1} \delta R) -  \lambda_{1} \delta R \exp(-\lambda_{1} \delta R) \bigr] \Bigr) \\
               = E \Bigl[ \exp( - \lambda_{1} (t-\delta) R)  - \exp(-\lambda_{1} t R) -  
                 \lambda_{1} \delta R \exp(-\lambda_{1} t R) \Bigr] .
    \end{multline*}
Therefore, by \eqref{E:probs.in.terms.of.f}, 
the Mean Value Theorem, and \eqref{E:f.beta.deriv}, 
 for some $\delta' \in [0, delta]$, we have
    \begin{equation*}  \label{E:M^delta.0.M_delta.>1}
         Prob \{ M^{\delta} = 0, \,  M_{\delta} > 1 \}  = - \delta f_{0}'(t-\delta') 
           - \lambda_{1} \delta f_{1}(t) 
             = \lambda \delta \bigl[ f_{1}(t - \delta') - f_{1}(t)  \bigr].
    \end{equation*}
Combining this with \eqref{E:M^delta+M_delta.decomp} and \eqref{E:M^delta>0.M_delta>0} we get
    \begin{multline} \label{E:M^delta+M_delta.prop.delta}
          Prob \{ M^{\delta} + M_{\delta} > 1, \, M_{\delta} > 0 \}  \\
            = \delta \lambda_{1} \Bigl(\bigl[ f_{1}(\epsilon')  -  f_{1}(t - \epsilon'') \bigr] 
              + \bigl[ f_{1}(t - \delta') - f_{1}(t)  \bigr] \Bigr) .
    \end{multline}

Next, in a similar way compute $Prob \{ M_{\delta} > 0 \}$. By \eqref{E:probs.in.terms.of.f} and \eqref{E:f.beta.deriv} there exists $\delta'' \in [0, \delta]$ s.t.\ 
    \begin{equation} \label{E:prob.M_delta.>0}
         Prob \{ M_{\delta} > 0 \} = 1 - f_{0}(\delta) = - \delta f_{0}'(\delta'') = \lambda_{1} \delta f_{1}(\delta'') .
    \end{equation}
Therefore, letting $\delta \downarrow 0$, we get finally,
    \begin{equation*} 
          Prob \{ M^{0^{+}} + M_{0^{+}} > 1 | M_{0^{+}} > 0 \}  \\
            = \frac{\bigl[ f_{1}(0)  -  f_{1}(t) \bigr] 
              + 0 }{ f_{1}(0) } = 1 - f_{1}(t)  .
    \end{equation*}
 
Next, let $N$ be the number of points that, given $R$, a Poisson process with rate $\lambda_{2} R$ puts in an interval disjoint from $[0,t]$ of length $T > 0$. 
    \begin{multline*}
         Prob \{ N = 0, \,  M_{\delta} > 0 \}  \\
           = E \Bigl( \exp( - \lambda_{2} T R) 
             \bigl[ 1 - \exp(-\lambda_{1} \delta R) \bigr] \Bigr) \\
               = E \Bigl[ \exp( - \lambda_{2} T R)   
                 - \exp \bigl( -(\lambda_{1} \delta + \lambda_{2} T) R \bigr) \Bigr] .
    \end{multline*}
    
 Define
    \begin{equation} \label{E:g.beta.defn}
      g_{\beta}(s) := \left( \frac{\alpha}{\alpha + \lambda_{1} s + \lambda_{2} T } \right)
        ^{\alpha+\beta}, \quad s > - (\alpha + \lambda_{2} T)/\lambda_{1} < 0  .
    \end{equation}
Thus,in general $g_{\beta}(0) \neq 1$ and 
    \begin{multline} \label{E:g.beta.deriv}
      g_{\beta}'(s) := - \lambda_{1} \frac{\alpha+\beta}{\alpha}
        \left( \frac{\alpha}{\alpha + \lambda_{1} s + \lambda_{2} T } \right)^{\alpha+\beta+1}
          =   - \lambda_{1} \frac{\alpha+\beta}{\alpha} g_{\beta+1}(s), \\
            s > - (\alpha + \lambda_{2} T)/\lambda_{1} < 0  .
    \end{multline} 
Then, by \eqref{E:general.R^q.exp(-lambda.R).formla}, and the Mean Value Theorem, there exists 
$\delta' \in [0, \delta]$ s.t.\
    \begin{equation} \label{E:g.beta.defn}
      Prob \{ N = 0, \,  M_{\delta} > 0 \} = g_{0}(0) - g_{0}(\delta)  = - \delta g_{0}'(\delta')
        = \delta \lambda_{1} g_{1}(\delta') .
    \end{equation}
 Hence, by \eqref{E:prob.M_delta.>0}, 
     \begin{equation} \label{E:g.beta.defn}
      Prob \{ N = 0 | M_{\delta} > 0 \} = g_{1}(\delta')/f_{1}(\delta'') .
    \end{equation}
 Letting $\delta \downarrow 0$, we get finally,
      \begin{equation} \label{E:g.beta.defn}
      Prob \{ N > 0 | M_{\delta} > 0 \} 
        1 - Prob \{ N = 0 | M_{\delta} > 0 \} = 1 - g_{1}(0) .
    \end{equation}

\subsubsection{Conditioning on zero baseline events} \label{SSS:zero.baseline}
Wasserman \emph{et al} \cite{dWcwHcWmW2015.SchoolBasedSuicidePrevention} has two special wrinkles. First, \emph{prima facie} the observations are not independent because students are randomized by school. However,  on p. 1541 we read ``Intraclass correlation across schools at 12 months were 0.003 for suicide SHB and 0.007 for severe suicide ideation.'' We will treat the observations as independent among subjects. In fact, not just for Wasserman \emph{et al} \cite{dWcwHcWmW2015.SchoolBasedSuicidePrevention}, but for all the studies \emph{we assume the students behave independently with respect to the variables of interest.}

The second wrinkle is not so easily dealt with. We are given, thank goodness, baseline counts of SHB (Table 1, p.\ 1538), but the follow-up counts only apply to those subjects who had no SHBs at baseline (Table 4, p.\ 1541). Let $u_{2} := Prob \{ N_{2} > 0 | N_{1} = 0 \}$. Then, by \eqref{E:unconditional.prob.of.0} or \eqref{E:qi.formla}, 
    \begin{align*}
        Prob \{ N_{2} > 0 \text{ and } N_{1} = 0 \} 
          &= E \bigl[ Prob \{ N_{2} > 0 \text{ and } N_{1} = 0 | R \} \bigr] \\
          &= E \Bigl( \bigl[ 1 - \exp(-R \lambda_{2} T_{2}) \bigr] \exp(-R \lambda_{1} T_{1}) \Bigr) \\
          &= E \Bigl( \exp(-R \lambda_{1} T_{1}) 
            - \exp \bigl[ -R (\lambda_{1} T_{1} + \lambda_{2} T_{2}) \bigr] \Bigr) \\
          &= \left( \frac{\alpha}{\alpha + \lambda_{1} T_{1}} \right)^{\alpha}  
            - \left( \frac{\alpha}{\alpha + \lambda_{1} T_{1}  + \lambda_{2} T_{2}} \right)^{\alpha}.
    \end{align*}
\eqref{E:unconditional.prob.of.0} tells us what $Prob \{ N_{1} = 0 \}$ is. Dividing the preceding expression by that probability we get, 
    \begin{equation}   \label{E:u2}
        u_{2} = 1 - \left( \frac{\alpha 
          + \lambda_{1} T_{1}}{\alpha + \lambda_{1} T_{1}  + \lambda_{2} T_{2}} \right)^{\alpha} .
    \end{equation}
A similar formula can be used for King \emph{et al} \cite{caKetal2018.suicidePrevention}. (See section \ref{SSS:Hazell} for the source of the baseline number for King \emph{et al} \cite{caKetal2018.suicidePrevention}.)

Fortunately, we know how many subjects had SHBs at baseline. So we can estimate $u_{2}$. Call that estimate $\hat{u}_{2}$. Setting the right hand side of \eqref{E:u2} equal to $\hat{u}_{2}$ gives us one of the equations we can use to estimate the parameters $\alpha$, $\lambda_{1}$ 
and the two $\lambda_{2}$'s one for each group. (See \eqref{E:4.eqns.in.4.unkns}, where different symbols are used for some of the parameters.) Plus, there's the equation for $r$, \emph{viz.}\ \eqref{E:gamma.mixture.correlation}. 
 
\subsubsection{McCauley}  \label{SSS:McCauley}
In \cite{eMcMEtal.2018.DBTforSuicidalBehavior} the inclusion criteria include that the subject must have made at least one suicide SHB. So we need some other fact about baseline SHB. In table 3, p.\ 782, we get data on SHBs in the last six months. Use those 6 month numbers. Thus, I'm interested in the conditional probability that someone has made an SHB in the 6 months prior to baseline given that they've made at least one SHB lifetime. There doesn't seem to have been any minimum age exclusion, but based on table 1, it seems extremely unlikely that the time, $\Delta$, from the age of risk (10 years) until recruitment age will be no more than $\delta := 6$ months. We assume that $\Delta > \delta$. Let $N$ be the number of SHBs a subject makes in the $\Delta$ years after entering the age of risk. Let $M$ be the number of SHBs in the $\delta$ years before enrollment. Condition on $\Delta$. Then I'm interested in 
    \begin{multline*}
      Prob \{ M > 0 | N > 0 \} = \frac{ Prob \{ M > 0 \text{ and } N > 0 \} }{ Prob \{ N > 0 \} } \\
        = \frac{ Prob \{ M > 0 \} }{ Prob \{ N > 0 \} } 
          = \frac{ 1 - E \exp( - R \lambda_{1} \delta ) }{ 1 - E \exp( - R \lambda_{1} \Delta ) } .
    \end{multline*}
Then as in \eqref{E:qi.formla}, 
    \begin{equation*}
      Prob \{ M > 0 | N > 0 \} 
        = \frac{ 1 - \left( \frac{\alpha}{\alpha + \lambda_{1} \delta} \right)^{\alpha} }
          { 1 - \left( \frac{\alpha}{\alpha + \lambda_{1} \Delta} \right)^{\alpha} } .
    \end{equation*}
This needs to be averaged over $\Delta$, which will follow a kind of truncated Gaussian distribution as in section \ref{SSS:truncated.normal}.

As for the follow-up probability, use formula \eqref{E:poz.in.flup.given.poz.at.base}. 

\subsubsection{Hazell} \label{SSS:Hazell}
On p. 663 of Hazell \emph{et al} \cite{plHgMkMcGtKaWgTrH2009.GroupTherapy}, under ``Entry criteria'' we read, ``Participants were eligible if they were aged between 12 and 16 years, had been referred to a child and adolescent mental health service in Australian sites \ldots and reported at least two episodes of self-harm in the past year, one of which had occurred in the past 3 months. Complicated!

To add to our troubles this paper doesn't seem to provide any data giving clues about the baseline population from which the sample is drawn, some mean or proportion that gives information about baseline SHB besides that given in the inclusion/exclusion criteria. For example in (McCauley, section \ref{SSS:McCauley} above) we used the proportion of subjects who had at least one SHB in the last 6 months prior to enrollment, an event not covered in the inclusion/exclusion in that paper. Another example is provided by Rossouw and Fonagy \cite{tiRpF2012.MentalizationTreatment} (section \ref{SS:Rossouw} below) where we used the proportion who first self-harmed in the last three months prior to enrollment, again something not covered in the inclusion/exclusion in that paper. But in Hazell \emph{et al} \cite{plHgMkMcGtKaWgTrH2009.GroupTherapy} and Ougrin \emph{et al} 
\cite{dOiBdSrBeT2013.SuicidePrevention} (subsubsection \ref{SS:prob.given.at.least.1}), I found nothing like that. 

To get around this problem I resorted to assuming a default rate of SHB, $\lambda_{1} = 0.023$ events per year. This number is derived from the estimate of the proportion of people who make an SHB during adolescence, which we took to be the period from age 10 to 18. That estimate is 16.1\% and comes from Muehlenkamp \emph{et al} \cite{jjMlClHplP2012.PrevalenceSHB}. That translates to a rate of 0.022 SHB per year.

Similarly, in King \emph{et al} \cite{caKetal2018.suicidePrevention} (subsubsection \ref{SSS:zero.baseline}) a useful baseline attempt rate is not present. Instead, I used an annual rated of about 0.005 attempts per years. That rate comes from the estimate of prevalence of 4.1\% in Nock \emph{el al} \cite{mkNjgGiHkaMcLnaSamZrcK2013.PrevalenceSA}. 

Let $\Delta$ be fixed at 1/4 year, i.e., 3 months. Let $t > 0$ be one year.
Let $M_{1}$ be the number of events in the interval $(0, t - \Delta)$ and let $M_{2}$ be the number of events 
in the interval $[t - \Delta, t)$. Let $N$ be the number of events in the fixed intervention period of length $T > 0$, a period disjoint from pre-baseline.  
We compute $Prob \{ N > 0 | M_{1} + M_{2} > 1, M_{2} > 0 \}$. We have
    \begin{multline*}
      Prob \{ M_{1} + M_{2} > 1, M_{2} > 0 \} = Prob \{ M_{1}  > 0, M_{2} > 0 \} 
         + Prob \{ M_{1} = 0, M_{2} > 1 \} \\
        = E \Bigl[ \bigl(1 
          - \exp( - \lambda_{1} R (t - \Delta) ) \bigr) \bigl(1 - \exp( - \lambda_{1} R \Delta ) \bigr) \Bigr] \\
          + E \Bigl[ \exp( - \lambda_{1} R (t - \Delta) ) 
            \bigl(1 - \exp( - \lambda_{1} R \Delta ) 
              - \lambda_{1} R \Delta \exp( - \lambda_{1} R \Delta )  \bigr) \Bigr] \\
              = 1 - E \exp( - \lambda_{1} R \Delta ) 
                - E \bigl[ \lambda_{1} R \Delta \exp( - \lambda_{1} R t ) \bigr] .
    \end{multline*}
Thus, by \eqref{E:qi.formla} and \eqref{E:unconditional.prob.of.1}, we get
    \begin{equation} \label{E:M1+M2>1.M2>0}
      Prob \{ M_{1} + M_{2} > 1, M_{2} > 0 \}  = 1 -  \left( \frac{\alpha}{\alpha + \lambda_{1} \Delta} \right)^{\alpha}
        - \lambda_{1} \Delta \left( \frac{\alpha}{\alpha + \lambda_{1} t } \right)^{\alpha+1} .
    \end{equation}

Similarly, 
    \begin{multline*}
      Prob \{ N > 0, M_{1} + M_{2} > 1, M_{2} > 0 \}  \\
      \begin{aligned}
         &= Prob \{ N > 0, M_{1}  > 0, M_{2} > 0 \} + Prob \{ N > 0, M_{1} = 0, M_{2} > 1 \}  \\
         &= E \Bigl[ \bigl(1 - \exp( - \lambda_{2} R T ) \bigr) 
           \bigl(1 - \exp( - \lambda_{1} R (t - \Delta) ) \bigr) 
            \bigl(1 - \exp( - \lambda_{1} R \Delta ) \bigr) \Bigr] \\
         & \qquad \qquad + E \Bigl[ \bigl(1 - \exp( - \lambda_{2} R T ) \exp( - \lambda_{1} R (t - \Delta) ) \\
         & \qquad \qquad \qquad \qquad \bigl(1 - \exp( - \lambda_{1} R \Delta ) 
            - \lambda_{1} R \Delta \exp( - \lambda_{1} R \Delta )  \bigr) \Bigr] \\
         &= 1 - E \exp( - \lambda_{1} R \Delta ) 
           - E \bigl[ \lambda_{1}  \Delta R \exp( - \lambda_{1} R t ) \bigr]  
            - E \exp( - \lambda_{2} R T ) \\
         & \qquad \qquad + E \exp( - R (\lambda_{1} \Delta + \lambda_{2} T) )
           + E \bigl[ \lambda_{1} \Delta R \exp( - R (\lambda_{1} t + \lambda_{2} T) ) \bigr] .
       \end{aligned}
    \end{multline*}
Therefore, by \eqref{E:unconditional.prob.of.0} and \eqref ,
    \begin{multline} \label{E:M1+M2>1.M2>0.N>0}
      Prob \{ M_{1} + M_{2} > 1, M_{2} > 0 , N > 0\}  \\
        = 1 -  \left( \frac{\alpha}{\alpha + \lambda_{1} \Delta} \right)^{\alpha}
        - \lambda_{1} \Delta \left( \frac{\alpha}{\alpha + \lambda_{1} t} \right)^{\alpha+1} 
          - \left( \frac{\alpha}{\alpha + \lambda_{2} T} \right)^{\alpha} \\
            + \left( \frac{\alpha}{\alpha + \lambda_{1} \Delta + \lambda_{2} T } \right)^{\alpha} 
             + \lambda_{1} \Delta \left( \frac{\alpha}{\alpha + \lambda_{1} t 
               + \lambda_{2} T} \right)^{\alpha+1} .
    \end{multline}

Thus, $Prob \{ N > 0 | M_{1} + M_{2} > 1, M_{2} > 0 \}$ is just \eqref{E:M1+M2>1.M2>0.N>0} divided by \eqref{E:M1+M2>1.M2>0}. 

\subsubsection{Mehlum}  \label{S:Mehlum}
The framework in section \ref{SSS:Hazell} is also used in Mehlum \emph{et al} \cite{lMetal2014.DialecticalBehaviorForSuicidalAdolescents}, except now $t$ is the number of years since the child turned 10, the age we take to be the beginning to the era of risk, and $\Delta$ is 16 weeks. 

Instead of giving numbers of subjects exhibiting SHB in one or more baseline periods, the paper gives summaries of subject level data. Now, mean number of SHB's is a more informative and easier to work with summary of subject behavior. Unfortunately, Mehlum \emph{et al} \cite{lMetal2014.DialecticalBehaviorForSuicidalAdolescents} doesn't use the mean at baseline. Instead, in Table 1, p.\ 1086 we get the \emph{median and interquartile range (IQR)} of the numbers of attempts lifetime. So we have to find values of $\alpha$ and $\lambda_{1}$ that generate per subject counts whose median and IQR best match the numbers in the table, \emph{conditioning} on the event $M_{1} + M_{2} > 1, M_{2} > 0$. (!) 

To do this we have to compute the conditional distribution of $M_{1}+M_{2}$ given the event
$M_{1} + M_{2} > 1$ and $M_{2} > 0$. Let $q$ be an integer greater than 1. (Given
$M_{1} + M_{2} > 1$ and $M_{2} > 0$, the sum $M_{1}+M_{2}$ has to be at least 2.)
We calculate $Prob \{ M_{1}+M_{2} = q | M_{1} + M_{2} > 1, M_{2} > 0 \}$. We have
    \begin{align*} 
      Prob \{ M_{1}+&M_{2} = q, M_{1} + M_{2} > 1, M_{2} > 0 \}  
          = Prob \{ M_{1}+M_{2} = q, M_{2} > 0 \} \\
        &= Prob \{ M_{1}+M_{2} = q \} - Prob \{ M_{1} = q, M_{2} = 0 \}  .
    \end{align*}
    \begin{align*} 
      Prob \{ M_{1} = q, M_{2} = 0 \} 
        &= E \left( \frac{\bigl[ \lambda_{1} R (t-\Delta) \bigr]^{q}}{q!} 
          \exp \bigl(-\lambda_{1} R (t-\Delta) \bigr) 
                \exp(-\lambda_{1} R \Delta) \right) \\
         &= \frac{\bigl[ \lambda_{1} (t-\Delta) \bigr]^{q}}{q!} 
           E \bigl[ R^{q} \exp(-\lambda_{1} R t) \bigr] .
    \end{align*}
Arguing as in \eqref{E:unconditional.prob.of.1} and \eqref{E:unconditional.prob.of.2} we see 
    \begin{equation}  \label{E:general.R^q.exp(-lambda.R).formla}
        E \bigl[ R^{q} \exp(-\lambda_{1} R t) \bigr] 
          =  \frac{\alpha^{\alpha}}{ (\alpha + \lambda_{1} t)^{\alpha+q}} \alpha (\alpha+1) 
            \cdots (\alpha+q-1) , \qquad q = 1, 2, \ldots .
    \end{equation}
(This formula actually generalizes appropriately to the $q = 0$ case. See \eqref{E:qi.formla} and \eqref{E:unconditional.prob.of.0}.)

Next, recall that the number of events in follow-up is, conditional on $R$, Poisson 
with rate $\lambda R T$. Compute $E ( N | M_{1} + M_{2} > 1, M_{2} > 0 )$. 
By \eqref{E:M1+M2>1.M2>0}, it suffices to compute 
$E ( N 1_{\{M_{1} + M_{2} > 1} 1_{\{M_{2} > 0\}} )$. Arguing as above, we have
    \begin{align*}
      E ( N & 1_{\{ M_{1} + M_{2} > 1\} } 1_{ \{ M_{2} > 0\} } ) \\
        &= E ( N 1_{\{ M_{1} + M_{2} > 1\} } ) 
          - E ( N 1_{\{ M_{1} > 1 \} } 1_{\{ M_{2} = 0\} } ) \\
        &= E \Bigl[ \lambda_{2} R T \,  \bigl( 1 - \exp (- \lambda_{1} R t) 
          - \lambda_{1} R t \exp (- \lambda_{1} R t) \bigr) \Bigr] \\
        & \qquad- E \Bigl( (\lambda_{2} R T)  \bigl[ 1 - \exp (- \lambda_{1} R (t-\Delta))  \\
        & \qquad \qquad  - \lambda_{1} R (t-\Delta) \exp (- \lambda_{1} R (t-\Delta)) \bigr] 
           \exp (- \lambda_{1} R \Delta)   \Bigr) .
    \end{align*}
Now, $E R=1$, by assumption. That, plus \eqref{E:general.R^q.exp(-lambda.R).formla} imply
    \begin{align*}
     \frac{1}{\lambda_{2} T}  E ( N & 1_{\{ \{M_{1} + M_{2} > 1\} } 1_{\{ \{M_{2} > 0\}| } ) \\
        &= E \biggl( R - R \exp(-\lambda_{1} R \Delta)- \lambda_{1} R^{2} \Delta \exp(-\lambda_{1} R t) \biggr) \\
        &= 1 - \left( \frac{\alpha}{ \alpha + \lambda_{1} \Delta} \right)^{\alpha+1}
           - \lambda_{1} \Delta \left( \frac{\alpha}{ \alpha + \lambda_{1} t} \right)^{\alpha+1} 
               \frac{\alpha+1 }{ \alpha + \lambda_{1} t} .
    \end{align*}
(See also \eqref{E:unconditional.prob.of.1} and \eqref{E:unconditional.prob.of.2}.) To compute the conditional expected value, just divide \linebreak
$E ( N 1_{\{ \{M_{1} + M_{2} > 1\} } 1_{\{ \{M_{2} > 0\}| } )$ by \eqref{E:M1+M2>1.M2>0}. \emph{For the rate per unit time divide the conditional expected value by $T$.}

\subsubsection{Rossouw and Fonagy} \label{SS:Rossouw}
In Rossouw and Fonagy \cite{tiRpF2012.MentalizationTreatment}, subjects must have committed self-harm in the last month before baseline. We know what proportion had first self-harmed in the last 3 months before baseline. 
Let $t > \Delta_{1} > \Delta_{2} > 0$. $t$ is the number of years since the subject entered the period of risk (i.e., years after 10 years of age), $\Delta_{1}$ is three months, and $\Delta_{2}$ is one month. Let $M_{1}$ be the number of events a subject had in the period $[0, t - \Delta_{1})$, let $M_{2}$ be the number in the interval $[t - \Delta_{2}, t]$. 
Let $\lambda_{1}$ be the Poisson rate in $[0, t]$. Let $T > 0$ and let $N$ be the number of events in an interval of length $T$ disjoint from $[0,t]$. Let $\lambda_{2}$ be the Poisson rate of occurrence in that rate. table 1, p.\ 1308 tells how many first harmed themselves within the last three months, i.e., within the period $t-\Delta_{1}$ to $t$. I'll take that as the baseline event. Thus, I'm interested in the conditional probabilities $Prob \{M_{1} = 0 | M_{2} > 0 \}$ and $Prob \{N > 0 | M_{2} > 0 \}$.
    
From \eqref{E:unconditional.prob.of.0} or \eqref{E:qi.formla},
    \begin{multline*}
      Prob \{M_{1} = 0, M_{2} > 0 \} 
        = E \left[  \exp \bigl( - \lambda_{1} R (t - \Delta_{1}) \bigr) 
           \bigl(1 - \exp ( - \lambda_{1} R \Delta_{2} ) \bigr) \right] \\
          = E \Bigl[ \exp \bigl( - \lambda_{1} R (t - \Delta_{1}) 
            - \exp \bigl( - \lambda_{1} R \bigl[ (t - \Delta_{1}) + \Delta_{2} \bigr] \bigr) \Bigr] \\
              = \left( \frac{\alpha}{\alpha + \lambda_{1} (t - \Delta_{1}) } \right)^{\alpha} 
                -\left( \frac{\alpha}{\alpha 
                  + \lambda_{1} \bigl[ (t - \Delta_{1}) + \Delta_{2} \bigr] } \right)^{\alpha} .
    \end{multline*}

    By \eqref{E:unconditional.prob.of.0} again, 
    \begin{multline*}
      Prob \{N > 0,  M_{2} > 0 \}
      = E \Bigl( \bigl[ 1 - \exp(-\lambda_{2} R T) \bigr] 
        \bigl[ 1 - \exp(-\lambda_{1} R \Delta_{2}) \bigr] \Bigr) \\
          = E \bigl[ 1 - \exp(-\lambda_{1} R \Delta_{2}) - \exp(-\lambda_{2} R T) 
            + \exp(-R (\lambda_{1} \Delta_{2} + \lambda_{2} T) \bigr] \\
              = 1 - \left( \frac{\alpha}{\alpha + \lambda_{1} \Delta_{2}} \right)^{\alpha}
                -  \left( \frac{\alpha}{\alpha + \lambda_{2} T} \right)^{\alpha} 
                  + \left( \frac{\alpha}{\alpha+ \lambda_{1} \Delta_{2} + \lambda_{2} T} \right)^{\alpha}  .
    \end{multline*}
To get the conditional probabilities divide by
    \begin{equation*}
         Prob \{ M_{2} > 0 \} = 1 - \left( \frac{\alpha}{\alpha + \lambda_{1} \Delta_{2}} \right)^{\alpha} 
    \end{equation*}
as before.

\subsection{Ideation} \label{S:ideation}

\subsubsection{Binary ideation variable}
In Wasserman \emph{et al} \cite{dWcwHcWmW2015.SchoolBasedSuicidePrevention}, suicidal observation is scored as a dichotomous variable. We can apply the methods described in section \ref{S:behavior}, especially subsection \ref{SSS:zero.baseline}, to analyze it. 

\subsubsection{Quantitative ideation measures}
In subsection \ref{SSS:gamma.mixture}, we introduced a multiplicative latent variable for modeling SHB. We do the same for quantitative ideation measures. Use an idea philosophically similar to that we use for dichotomous data and described in subsection \ref{SSS:gamma.mixture}, viz., using a latent variable to give rise to correlation, but we use a lognormal mixture, not a gamma. Let $X_{pre}$ and $X_{post}$ be ideation measurements for one arm of the trial. Suppose they have the form
    \begin{equation} \label{E:X.pre.and.X.post.lognormal.defn}
        X_{pre} = \exp(Z + W_{pre}) \text{ and } X_{post} = \exp(Z + W_{post}),
    \end{equation}
where $Z \approx N(0, \sigma^{2})$, $W_{pre} \approx N(\mu_{pre}, \tau_{pre}^{2})$, and 
$W_{post} \approx N(\mu_{post}, \tau_{post}^{2})$ are independent and normal. Thus $X_{pre}$, $X_{post}$ are lognormal. 

From Taylor and Karlin \cite[p.\ 28]{hmTsK97.StochasticModeling} or Wikipedia we have
    \begin{multline} \label{E:lognormal.mean.variance}
        \text{mean of } X_{pre} 
          = E X_{pre} = \exp \left[ \mu_{pre} + \tfrac{1}{2}(\sigma^{2} + \tau_{pre}^{2}) \bigr] \right] 
            \text{ and } \\
          Var X_{pre} = \exp \bigl[ 2 \mu_{pre} + (\sigma^{2} + \tau_{pre}^{2}) \bigr] 
            \bigl[ \exp (\sigma^{2} + \tau_{pre}^{2}) - 1 \bigr] .
    \end{multline}
Similarly for $X_{post}$.

Notice that $X_{pre} X_{post} = \exp ( 2Z + W_{pre} + W_{post})$ and hence $X_{pre} X_{post}$ is also lognormal. Therefore, 
    \begin{equation*}
        E (X_{pre} X_{post})  
          = \exp \left[ \mu_{pre} + \mu_{post} + 2 \sigma^{2} + \tfrac{1}{2}(\tau_{pre}^{2} 
            + \tau_{post}^{2}) \right] .
    \end{equation*}
The pre-post correlation can be computed from this and \eqref{E:lognormal.mean.variance}.

If there are two follow-up groups we end up with seven parameters to be estimated from the data.

\subsubsection{Diamond ideation} \label{SSS:Diamond.ideation}
In Diamond \emph{et al} \cite{gsDetAl.2010.attachment.family} and \cite{gsDrrKesKEsaLjlHjmRrjG2019.attachment}, subjects enrolled in the study had to have a minimum ideation score. Here we develop a formula for computing moments conditional on this requirement. Let $x \in \RR$ and let $p, q = 0, 1$ or 2. Use the notation above. We're interested in $E \bigl[ X_{pre}^{p} X_{post}^{q} | X_{pre} > x \bigr]$ because if we can  compute that expectation then we can compute means ($p=1$, $q=0$ \emph{or} $p=0 $, $q=1$) and second moments ($p=2$, $q=0$ \emph{or} $p=0$, $q=2$ \emph{or} $p=1$, $q=1$). This allows us to also compute variances and correlation. We have
    \begin{align*}     \label{E:restricted.p.q.moment}
      E \bigl[ X_{pre}^{p} &X_{post}^{q} 1_{\{X_{pre} > x\}} \bigr] \notag \\
        &= E \Bigl( E \bigl[ X_{pre}^{p} X_{post}^{q} 1_{\{X_{pre} > x\}} | Z \bigr] \Bigr) \notag \\
        &= E \Bigl( E \bigl[ e^{p W_{pre}} e^{q W_{post}} e^{(p+q)Z} 1_{\{X_{pre} > x\}} 
          | Z \bigr] \Bigr) .
     \end{align*} 
 Now, $W_{post}$ is independent of $W_{pre}$ and $Z$, so we can pull it out of the conditional expectation, yielding,   
     \begin{equation*}
     E \bigl[ X_{pre}^{p} X_{post}^{q} 1_{\{X_{pre} > x\}} \bigr]
       = E e^{q W_{post}} \;
           E \Bigl( E \bigl[ e^{p W_{pre}} e^{(p+q)Z} 1_{\{X_{pre} > x\}} | Z \bigr] \Bigr) .
    \end{equation*}
Since $e^{(p+q)Z}$ is a function of $Z$ it also can come out of the conditional expectation:
     \begin{equation*}
     E \bigl[ X_{pre}^{p} X_{post}^{q} 1_{\{X_{pre} > x\}} \bigr]
       = E e^{q W_{post}} \;  
         E \Bigl( \exp \bigl[ (p+q)Z \bigr] 
           E \bigl[ e^{p W_{pre}} 1_{\{X_{pre} > x\}} | Z \bigr] \Bigr).
    \end{equation*}
Use the obvious analogue of \eqref{E:lognormal.mean.variance} to compute $E \exp{ \bigl[ q W_{post} + (p+q)Z \bigr] }$. By \eqref{E:X.pre.and.X.post.lognormal.defn}, 
$1_{\{X_{pre} > x\}} = 1_{\{W_{pre} > \log{x} - Z\}}$. Since $W_{pre}$ is independent of $Z$, the conditional expectations can be written as a function of $Z$. Let 
    \begin{equation*}
        \zeta(z) := \zeta(z;x)
          :=  E \bigl[ e^{p W_{pre}} 1_{\{W_{pre} > \log x - z \}} \bigr] .
    \end{equation*}
Then $E \bigl[ e^{p W_{pre}} 1_{\{X_{pre} > x\}} | Z \bigr] = \zeta(Z)$. Thus, 
    \begin{equation}  \label{E:restricted.p.q.moment.simpfied} 
       E \bigl[ X_{pre}^{p} X_{post}^{q} 1_{\{X_{pre} > x\}} \bigr] 
         = \exp \Bigl( q \mu_{post} 
           + \tfrac{1}{2} \bigl[ (p+q)^{2} \sigma^{2} + q^{2} \tau_{post}^{2} \bigr] \Bigr) E \zeta(Z) .
    \end{equation}
The preceding quantity must now be divided by 
$Prob \{ X_{pre} > x \} = Prob \{ Z + W_{pre} > \log x \}$. Now, $Z + W_{pre}$ is normal 
$N(\mu_{pre}, \sigma^{2} + \tau_{pre}^{2})$. So $Prob \{ X_{pre} > x \}$ is a normal tail probability.

We compute $E \zeta(z)$ for arbitrary real $z$. Let $\varphi(\cdot; \mu, \tau)$ be the $N(\mu, \tau^{2})$ density. Then,
    \begin{equation*}
      \zeta(z) = \int_{\log x - z}^{\infty} \exp(p w) \, \varphi(w; \mu_{pre}, \tau_{pre}) \, dw .
    \end{equation*}
The integrand in the preceding is
    \begin{align}  \label{E:gaussian.density.with.exponential.factor}
    \exp(p w) \, \varphi(w;  &\mu_{pre}, \tau_{pre}) = \exp(p w) \frac{1}{\sqrt{2 \pi \tau_{pre}^{2}}} 
        \exp \left( - \frac{(w - \mu_{pre})^{2}}{2 \tau_{pre}^{2}} \right) \notag \\
          &= \frac{1}{\sqrt{2 \pi \tau_{pre}^{2}}} 
            \exp \left( p w - \frac{(w - \mu_{pre})^{2}}{2 \tau_{pre}^{2}} \right) \notag \\
          &= \frac{1}{\sqrt{2 \pi \tau_{pre}^{2}}} 
            \exp \left(- \frac{ - 2 p \tau_{pre}^{2} w + (w - \mu_{pre})^{2}}{2 \tau_{pre}^{2}} \right) \notag \\
          &= \frac{1}{\sqrt{2 \pi \tau_{pre}^{2}}} 
            \exp \left(- \frac{ - 2 p \tau_{pre}^{2} w + w^{2} - 2 w \mu_{pre} + \mu_{pre}^{2}}
              {2 \tau_{pre}^{2}} \right) \\
          &= \frac{1}{\sqrt{2 \pi \tau_{pre}^{2}}} 
            \exp \left(- \frac{ w^{2} - 2 (\mu_{pre} + p \tau_{pre}^{2}) w + \mu_{pre}^{2}}
              {2 \tau_{pre}^{2}} \right) \notag \\
          &= \frac{1}{\sqrt{2 \pi \tau_{pre}^{2}}} 
            \exp \left(- \frac{ \bigl[ w - (\mu_{pre} + p \tau_{pre}^{2}) \bigr]^{2} 
             - 2 \mu_{pre} p \tau_{pre}^{2} - p^{2} \tau_{pre}^{4} }
               {2 \tau_{pre}^{2}} \right) \notag \\
          &= \exp ( \mu_{pre} p + p^{2} \tau_{pre}^{2}/2 ) \, 
            \varphi(w; \mu_{pre} + p \tau_{pre}^{2}, \tau_{pre})  
            \notag .
    \end{align}
Let $\Phi(\cdot; \mu, \tau)$ be the $N(\mu, \tau^{2})$ cumulative distribution function. Then,
    \begin{multline*}
      \zeta(z) = E \bigl[ e^{p W_{pre}} 1_{\{W_{pre} > \log x - z \}} ] \\
        = \exp ( \mu_{pre} p + p^{2} \tau_{pre}^{2}/2 ) \, 
          \bigl[  1 - \Phi(\log x - z ;  \mu_{pre} + p \tau_{pre}^{2}, \tau_{pre}) \bigr] .
    \end{multline*}

We get
    \begin{equation*}
           \exp ( q \mu_{post} + \tfrac{1}{2}q^{2} \tau_{post}^{2} ) \; 
                  e^{(p+q)z} \zeta(z;x)
    \end{equation*}
The next step is to multiply $e^{(p+q)z} \zeta(z)$ by $\varphi(z; 0, \sigma)$ and integrate 
over $z \in \RR$. Finally, we divide the resulting value 
by $Prob \{ X_{pre} > x \} = Prob \{ Z + W_{pre} > \log x \}$, a normal probability. 

\section{Acknowledgements}
I wish to thank my co-authors on Itzhaky \emph{et al}  \cite{lIetAl2021.SuicidePreventionMetaAnalysis}: 
                          Liat Itzhaky,
                          Sara Davasaambuu, 
        			Sebastian Cisneros-Trujillo, 
        			Katrina Hannett, 
        			Kelly Scolaro, 
        			Jamil Lemberansky, 
        			Barbara Stanley, 
        			J. John Mann, 
        			Milton L. Wainberg, 
        			Maria A. Oquendo,  and 
        			M. Elizabeth Sublette .

\newcommand{\etalchar}[1]{$^{#1}$}
\providecommand{\bysame}{\leavevmode\hbox to3em{\hrulefill}\thinspace}
\providecommand{\MR}{\relax\ifhmode\unskip\space\fi MR }
\providecommand{\MRhref}[2]{%
  \href{http://www.ams.org/mathscinet-getitem?mr=#1}{#2}
}
\providecommand{\href}[2]{#2}

\end{document}